\pdfoutput=1
\documentclass[11pt]{article}

\usepackage[a4paper, margin=1in]{geometry}
\usepackage[T1]{fontenc}
\usepackage{lmodern}
\usepackage{microtype}
\usepackage{amsmath, amssymb}
\usepackage{mathrsfs}
\usepackage{graphicx}
\usepackage{xcolor}
\usepackage{booktabs}
\usepackage{colortbl}
\usepackage{caption}
\usepackage{authblk}
\usepackage[numbers, sort]{natbib}
\usepackage[section]{placeins}
\usepackage{hyperref}
\hypersetup{colorlinks=true,linkcolor=blue!50!black,
            citecolor=blue!50!black,urlcolor=blue!50!black}

\newcommand{\lf}{\lambda_F}%
\newcommand{\Me}{M_e}%
\newcommand{\TM}{T_M}%
\newcommand{\TF}{T_F}%
\newcommand{\Hfield}{H}%
\newcommand{\hwave}{h}%
\newcommand{\hzero}{h_0}%
\newcommand{\hs}{h_s}%
\newcommand{\Xp}{\boldsymbol{X}_p}%
\newcommand{\Pback}{P_{\mathrm{back}}}
\newcommand{\Pfwd}{P_{\mathrm{fwd}}}
\newcommand{\floop}{f_{\mathrm{loop}}}
\renewcommand{\vec}[1]{\boldsymbol{#1}}

\title{\bfseries Looped trajectories in triple-slit diffraction\\[3pt]
       with pilot-wave hydrodynamics}

\author[1]{Georgi~Gary~Rozenman}
\author[1,*]{John~W.~M.~Bush}
\affil[1]{Department of Mathematics, Massachusetts Institute of Technology,
          Cambridge, MA 02139, USA}
\affil[*]{Correspondence: \texttt{bush@math.mit.edu}}
\date{29 November 2026}

\begin{document}
\maketitle

\begin{abstract}
\noindent
In pilot-wave hydrodynamics, a millimetric bouncing droplet is guided
across the surface of a vibrating liquid bath by its own wave field. As
this wave field encodes the history of the droplet's motion, one may
view the resulting wave force as arising from a self-potential
associated with the droplet's memory. Using an established theoretical
model of this system, we examine droplet diffraction through a
three-slit barrier and discover exotic loop trajectories of the form
predicted theoretically, but not directly observable, in quantum
diffraction.
With our chosen geometry, $0.80\,\%$ of the droplets that reach the screen loop back and cross the
barrier more than once en route, threading all three apertures in
turn.
The loop trajectories may be rationalized in
terms of the distortion of the droplet's pilot wave by the barrier, and the
resulting spiral scattering of the droplet. We
compare the rate of backward transport of droplets with that of a
wavepacket in the Schr\"{o}dinger equation, and find backward transport
of comparable magnitude in the two systems.
\end{abstract}

\section*{Introduction}

The pilot-wave hydrodynamic system consists of millimetric droplets of
silicone oil self-propelling on the surface of a vibrating bath of that
same liquid~\citep{Couder2005}. The resulting walking droplets are dressed
in a quasi-monochromatic wave field, the presence of which is
responsible for the system's myriad quantum
features~\citep{Bush2015AnnRev,BushOza2021,bush2024perspectives}. The
walkers comprise both droplet and guiding wave, and so represent a 
concrete, classical example of wave--particle
duality. Moreover, they have provided a macroscopic realization of a
pilot-wave system of the form envisaged by de~Broglie in his
double-solution pilot-wave theory~\citep{deBroglie1987}, according to
which quantum particles have an intrinsic vibration responsible for the
local generation of a spatially-extended pilot
wave~\citep{Bush2015AnnRev}. De~Broglie proposed, but never proved,
that his pilot-wave dynamics might give rise to statistical behaviour
consistent with that predicted by the standard quantum formalism, a
proposition
that has received some support from pilot-wave
hydrodynamics~\citep{CouderFort2006,Harris2014corral,Saenz2018Friedel}.

The walking-droplet system has captured several features previously
thought to be exclusive to the quantum realm, providing classical
analogues of single- and multiple-slit
diffraction~\citep{CouderFort2006,Pucci2018Faraday}, quantum
corrals~\citep{Harris2014corral,Saenz2018Friedel}, Friedel
oscillations~\citep{Saenz2019a},
tunnelling~\citep{Eddi2009Tunnelling} and quantized
orbits~\citep{Fort2010Quantization,Perrard2014,Harris2014a}.
More recently, it has provided hydrodynamic analogues of spin
lattices~\citep{Saenz2021SpinLattices}, Anderson
localization~\citep{Abraham2024}, interaction-free
measurement~\citep{FrumkinBush2023}, static Bell
violations~\citep{Papatryfonos2024} and the Aharonov--Bohm
effect~\citep{Rozenman2025AB}. Pilot-wave hydrodynamics has thus
launched the field of hydrodynamic quantum analogues
(HQAs)~\citep{BushOza2021}, and motivated the exploration of a
broader class of classical pilot-wave
theories~\citep{Durey2021,DarrowBush2024,JametDrezet2021} that 
are together serving to redefine the boundary between classical and 
quantum physics.

A characteristic feature of free walkers is that, in the bulk of
parameter space, they self-propel along straight lines at a constant
speed~\citep{Oza2013}. Deviations from rectilinear motion arise only
in response to disruptions, as may be imposed by applied forces, boundaries or other
droplets. Another notable feature of pilot-wave hydrodynamics is spin
states, in which droplets execute a circular orbit with radius
$\approx\lf/2$~\citep{Oza2018,Labousse2016}. While stationary spin
states are weakly unstable in the laboratory, one may view vestiges
of them when the drops move in response to confining
forces~\citep{Labousse2016} or bounding
geometry~\citep{Harris2014corral}. Drifting spin states were
reported by Bernard-Bernardet \emph{et
al.}~\citep{BernardBernardet2022}. The associated tendency of
walkers to turn on an arc with radius of curvature $R\sim\lf/2$ is
related to the spiral scattering reported by Harris {\it et al.} \citep{Harris2018}.
Specifically, when a walker encounters a submerged pillar, it is scattered
and departs along a logarithmic spiral, the radius of curvature of which
increases monotonically as it unwinds~\citep{Harris2018}. This
spiral scattering will be key to understanding the phenomenon reported here,
and will guide our selection of the slit geometry.

The field of hydrodynamic quantum analogues was initiated by Couder and
Fort's pioneering investigation of walker
diffraction~\citep{CouderFort2006}, in which walkers were launched
towards submerged topography that served as single and double slits,
prompting deflection of the walkers and the emergence of coherent
diffraction patterns from an ensemble of individual walkers.
The authors reported Fraunhofer-like diffraction patterns, and the
independence of the deflection angle on the \emph{impact parameter}
--- the offset of the walker's incoming path from the slit centreline
--- suggesting the dominance of chaos in the system. It is notable that the field of
pilot-wave hydrodynamics was in its infancy, so the experiments were
not as thoroughly reported as they might have been. In particular, the
authors reported neither drop size nor the vibrational acceleration of
the bath. It is now known that the system behaviour is extremely
sensitive to both quantities, which has made it extremely difficult
to reproduce their experimental results~\citep{Pucci2018Faraday}.

While early attempts to recreate coherent diffraction patterns were
unsuccessful~\citep{Andersen2015,Batelaan2016}, more controlled
experiments~\citep{Pucci2018Faraday,Ellegaard2020} have revealed
coherent diffraction patterns, but with two notable distinctions from the results of Couder
\& Fort. First, the trajectories were relatively predictable: chaotic
trajectories were relatively rare, arising only in the high-memory
limit when the resonance between drop and wave was
lost~\citep{Wind2013}. Elsewhere,
there is a direct map between impact parameter and deflection angle.
Second, the diffraction patterns did not conform to the Fraunhofer
pattern. While the shortcomings of Couder \& Fort's study have now been
aired, its two central conclusions have been confirmed. First,
coherent, repeatable statistical wave patterns emerge from
single-particle walker diffraction. Second, in the double-slit
geometry, the far-field diffraction pattern is altered relative to
that of the single slit by the presence of the second slit, owing to
its influence on the wave field and so the path of the droplet passing
through the first.

While the bulk of the studies of walker diffraction have focused on the
distribution of deflection angles in the far field, Pucci \emph{et
al.}~\citep{Pucci2025} examined the behaviour of the walkers in the
near field, specifically, within a few wavelengths of the slits. Using
the stroboscopic model of walker dynamics~\citep{Oza2013} and modelling
the slits as a linear array of wave sources~\citep{Couder2005}, they
reported the similarities between the walker behaviour and near-field
diffraction patterns of electrons and photons. Specifically, they
demonstrated that as the droplet crosses the slit, it excites a
transient wave form spanning the slit that prompts channelling across a
number of preferred crossing points separated by half a Faraday
wavelength. Notably, similar channelling has been reported in experiments on the
hydrodynamic analogue of Kapitza--Dirac diffraction, in which walkers
cross a standing field of Faraday waves~\citep{primkulov_KD_2025}. Channelling of this type also arises in a classical, relativistic
pilot-wave theory of single-particle diffraction that captured
Fraunhofer diffraction patterns from both single and double
slits~\citep{DarrowBush2024,DarrowBush2025}. We here focus on an
intriguing feature of near-field walker diffraction arising in the triple-slit
geometry.

Loop trajectories are predicted to arise in the quantum-mechanical
treatment of triple-slit diffraction~\citep{Sawant2014}. In the
Feynman path-integral
formulation, a particle reaches the screen along all possible paths,
including those that thread the three slits in turn
(Fig.~\ref{fig:diffraction}). Such
looping paths are associated with \emph{probability backflow}: even a
state built entirely from forward-moving (positive-momentum) components
can, for a time, carry probability backwards. Thus, the local
probability current at a point can oppose the packet's mean direction
of motion~\citep{Allcock1969,BrackenMelloy1994}. The probability of
such backflow is bounded by the Bracken--Melloy constant
$c_{\mathrm{bm}}\approx0.0384$~\citep{BrackenMelloy1994,Penz2006},
the largest fraction of particles that may flow backward for a
state composed purely of forward momenta.
In the quantum setting, these exotic loop trajectories cannot be
observed directly;
they are inferred only indirectly, from the emergent diffraction
pattern~\citep{Eliezer2020,Barbier2024Backflow}. They were the target
of the photonic three-slit experiment of Maga\~{n}a-Loaiza \emph{et
al.}~\citep{MaganaLoaiza2016}, who used surface-plasmon near fields to
enhance the prevalence of the looped paths. Their study built upon
the experiments of Sinha \emph{et al.}~\citep{Sinha2010}, who compared the interference patterns arising from multi-slit and
pairwise-slit geometries. From the far-field statistics, Maga\~{n}a-Loaiza \emph{et al.}
inferred that such loops occur in roughly one in a hundred
trajectories. We show here that loop trajectories of this kind, while
similarly rare, are directly
observable in pilot-wave hydrodynamics, where they arise through the
spiral-scattering mechanism~\citep{Harris2018} that takes place within
the slit array.

We first detail the physical system of interest in addition to the theoretical model used to describe it. We then describe the behaviour predicted by our theoretical model. Particular attention is given to characterizing the exotic loops and the factors influencing their likelihood.
We demonstrate that the exotic trajectories arise only for certain values of the impact parameter, the distance of the initial trajectory from the centreline of the slit array. Finally, we compare and contrast the observed behaviour with that reported in its quantum counterpart.
\section*{Physical picture and theoretical model}

When a liquid bath is vibrated vertically with frequency $\omega$ and
peak acceleration $\gamma$ exceeding the Faraday threshold
$\gamma_F$, its surface destabilizes into a field of standing,
subharmonic Faraday waves~\citep{Miles1990} with frequency
$\omega_F = \omega/2$ and wavelength $\lf = 2\pi/k_F$ prescribed by the
water-wave dispersion relation. For $\gamma < \gamma_F$, millimetric
droplets may bounce on the vibrating bath, their coalescence precluded
by the sustenance of a lubrication layer of air during
impact~\citep{Couder2005b,Molacek2013a}. When the droplet bounces at
the Faraday frequency, in resonance with the most unstable wave mode
of the bath, it may destabilize from a static bouncing state into a
self-propelling walking state. The resulting walking drop moves 
in response to the local slope of its wave field
$\Hfield(\boldsymbol{x},t)$~\citep{Molacek2013b,Oza2013,Pucci2025},
a superposition of spatially and temporally damped Bessel functions
$J_0(k_F r)$ generated at prior impacts along its
path~\citep{Molacek2013b}. The field thus carries a memory of the
walker's path, the extent of which is prescribed by the dimensionless
memory parameter $\Me = \TM/\TF$, where $\TM$ is the time over which a
wave on the bath decays and $\TF$ the Faraday period (Methods). As
$\Me$ increases monotonically with $\gamma/\gamma_F$, the vibrational
acceleration is used to prescribe the system memory.

We model the walker dynamics using the stroboscopic pilot-wave model
of Oza, Rosales \& Bush~\citep{Oza2013,Labousse2014}. The drop of
mass $m$ with horizontal position $\Xp(t)$ obeys the trajectory
equation
\begin{equation}
m\,\ddot{\Xp} + D\,\dot{\Xp}
  \;=\; -m g\,\nabla\Hfield(\Xp,\,t),
\label{eq:eom}
\end{equation}
where $D$ is the time-averaged horizontal drag~\citep{Oza2013} and the
pilot-wave field
\begin{equation}
\Hfield(\boldsymbol{x},\,t)
  \;=\; \frac{1}{\TF}\!\int_{-\infty}^{t}\!
       \hwave(\boldsymbol{x},\,s)\,
       e^{-(t-s)/\TM}\,\mathrm{d}s,
\label{eq:Hfield}
\end{equation}
is the field generated along the walker's path. For a free walker,
$\hwave(\boldsymbol{x},s) = \hzero\!\bigl(|\boldsymbol{x}-\Xp(s)|\bigr)$,
where $\hzero(r) = J_0(k_F r)\,e^{-r/\delta}$ describes the wave form
generated by each impact, a Bessel function damped over a lengthscale
$\delta$~\citep{Pucci2025}. Throughout our study, we take $\delta = 1.5\,\lf$
to be consistent with prior work~\citep{Damiano2016,Tadrist2018a}.
The boundary treatment follows Pucci \emph{et al.}~\citep{Pucci2025},
who modelled the barrier as an array of wave sources spaced a
distance $\lf/2$ apart (Methods and Supplementary Sec.~S3) that serve to locally
suppress the pilot wave. The total wave field
$\hwave(\boldsymbol{x},s)$ may then be decomposed into contributions 
from the walker $\hzero$ and the barrier sources $\hs$:
\begin{equation}
\hwave(\boldsymbol{x},\,s)
 \;=\; \hzero\!\bigl(|\boldsymbol{x}-\Xp(s)|\bigr)
 \;+\; \hs(\boldsymbol{x},\,s),
\qquad
\hs(\boldsymbol{x},\,s)
 \;=\; \sum_{j} W_j(s)\,
       \hzero\!\bigl(|\boldsymbol{x}-\boldsymbol{x}_j|\bigr).
\label{eq:hwave}
\end{equation}
The source weights $W_j(s)$ at the barrier locations
$\boldsymbol{x}_j$ are determined at every instant by enforcing
the Dirichlet condition such that $\hwave(\boldsymbol{x}_j,s) = 0$ at
the position of every source~\citep{Pucci2025}
(Fig.~\ref{fig:diffraction}a).
The trajectory equation~\eqref{eq:eom} is integrated by a Verlet
scheme, and the memory integral~\eqref{eq:Hfield} is evaluated as a
discrete exponentially weighted sum truncated after $5\,\Me$ steps,
an approximation that is equivalent to Eq.~\eqref{eq:Hfield} to 
numerical precision (Methods).

The system geometry is illustrated in Fig.~\ref{fig:diffraction}a. 
Three slits ($S_1, S_2, S_3$) of equal width $L = 1.2\,\lf$ are separated by barrier segments ($B_1, B_2$) of width $b = 0.3\,\lf$ and bounded by the outer walls ($W_1, W_2$). 
The barriers and walls are infinitely thin, represented by a single line of 
wave sources at $y=0$ (white dots in Fig.~\ref{fig:diffraction}a). Along each outer wall the sources are spaced $\lf/2$ apart, the first lying at the edge of the adjacent slit; each barrier segment, being narrower than this spacing, is represented by two sources, one at each of its edges, a distance $b=0.3\,\lf$ apart.
We consider the high-memory regime ($\Me = 50$), which requires evaluation of the last 250 impacts in the memory integral.
The choice of slit geometry was informed by the pillar-scattering study of Harris {\it et al.}~\citep{Harris2018}: with $L=1.2\,\lf$, adjacent slits lie within reach of a single spiral turn prompted by scattering from one of the barriers. Figure~\ref{fig:diffraction}b illustrates the looping phenomenon of interest: a walker launched towards the rightmost slit $S_3$ passes through all three slits before proceeding to the screen.

\section*{Triple-slit diffraction and looped trajectories}

\begin{figure}[tbp]
\centering
\includegraphics[width=\linewidth]{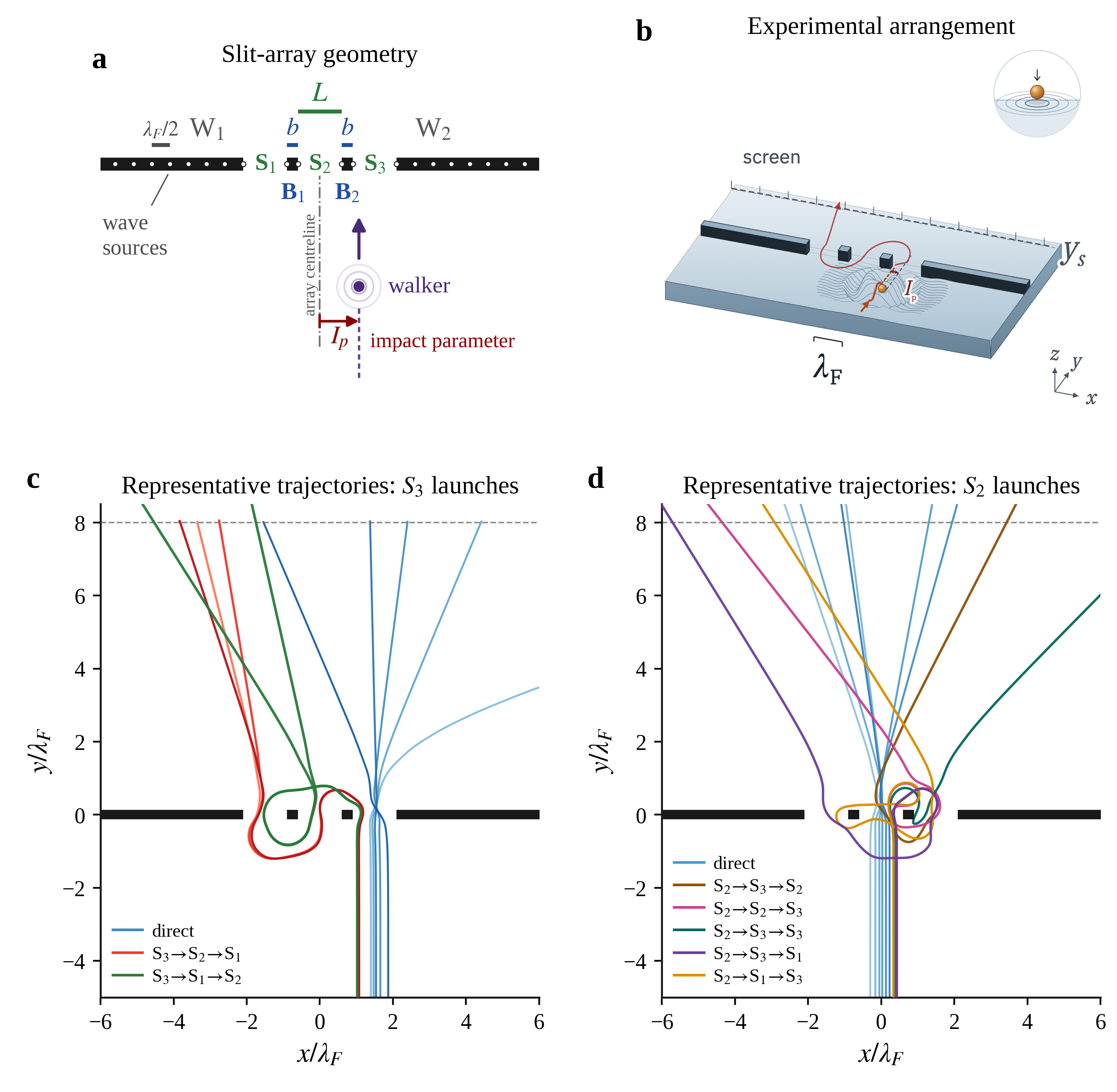}
\caption{\textbf{Exotic loop trajectories in triple-slit walker
diffraction.}
(a)~Geometry of the slit array: slits $S_1$, $S_2$ and $S_3$ of
width $L$ (green), separated by the interior barrier segments $B_1$ and $B_2$
of width $b$ (blue) and bounded by the outer walls $W_1$ and $W_2$. The
barrier is realized as a line of point wave sources (white dots): they are
spaced $\lf/2$ apart along each outer wall, beginning at the slit edge, and
each interior segment carries two, one at each of its edges (Methods).
A walker is launched towards
$S_3$ with impact parameter $I_p$, the distance of its initial incoming path
(dashed trail) to the right of the centreline of the slit array (dash-dot line).
(b) Schematic of the experimental arrangement: a walker (orange) launched towards $S_3$
rides its pilot wave towards the submerged three-slit barrier
and onwards to the screen (dashed line at $y_s$). 
(c)~Representative trajectories of walkers launched through $S_3$:
direct walkers (blue) and five loop trajectories 
that recross the barrier and thread all three apertures before 
reaching the screen ($y=8\,\lf$). Three follow the dominant
$S_3\!\to\!S_2\!\to\!S_1$ route (red) and two the
$S_3\!\to\!S_1\!\to\!S_2$ path (green).
(d)~Representative trajectories of walkers launched through $S_2$
(Supplementary Sec.~S2): direct walkers (blue) and one
example of each exotic loop that reaches the screen. These include
$S_2\!\to\!S_3\!\to\!S_2$, which executes a full clockwise orbit of $B_2$ (brown);
$S_2\!\to\!S_2\!\to\!S_3$ (magenta); $S_2\!\to\!S_3\!\to\!S_3$
(teal); the three-slit path $S_2\!\to\!S_3\!\to\!S_1$ (purple); and
$S_2\!\to\!S_1\!\to\!S_3$ (orange). Launches to the left of the 
centreline ($I_p<0$) are not reported as their forms are identical 
but mirror symmetric. Walkers that travel back towards the source are
not drawn; the largest such group, walkers that return through $S_3$
and cross no more, makes up $15\,\%$ of the $S_2$ launches. The
loops of both ensembles are overlaid in Supplementary
Fig.~\ref{fig:allloops}.
}
\label{fig:diffraction}
\end{figure}

Walking droplets are launched towards the triple-slit arrangement.
Each launch is specified by its impact parameter $I_p$, the position of the launch point relative to the centreline of the slit array
(Fig.~\ref{fig:diffraction}a). The system symmetry allows us to completely characterize its behaviour through consideration of $I_p \geq 0$, that is, of launches through the right half of the central slit $S_2$ and through the rightmost slit $S_3$: launches with $I_p<0$ follow the mirror-image trajectories. Our ensembles sample $I_p$ uniformly across the entire span of each slit, less a margin of $0.05\,\lf$ at their edges, specifically $0 \leq I_p \leq 0.55\,\lf$ for launches through $S_2$, and $0.95\,\lf \leq I_p \leq 2.05\,\lf$ for launches through $S_3$ (Methods). These margins are excluded because walkers launched within them never reach the screen; instead, they pass through the slit and are then absorbed by the barrier (Supplementary Sec.~S1). Most walkers cross the barrier only once and proceed
to the screen (Fig.~\ref{fig:diffraction}c, blue), but a small fraction execute the looping phenomenon of interest. 
For example, after passing through the rightmost slit $S_3$, they may turn 
back, then thread both of the other slits before reaching the screen 
(Fig.~\ref{fig:diffraction}c, red and green). Walkers launched through the central slit $S_2$ (Fig.~\ref{fig:diffraction}d) execute a wider variety of exotic itineraries, to which we return below. 
The loops of both ensembles are overlaid in Supplementary Fig.~\ref{fig:allloops}.

Figure~\ref{fig:wavefield} shows the evolution of the pilot wave as the droplet loops through the slits. The walker approaches the barrier $B_2$, its pilot wave distorted by the presence of the slits (a,b). Deflected as it crosses $S_3$, it locks onto a spiralling trajectory about $B_2$, which
returns it to $S_2$ (c). It then executes a second spiral about
$B_1$, from which it emerges through $S_1$ (d), departing towards the
screen as its pilot wave relaxes to that of a free walker (e,f). The
loop thus consists of a pair of spiral-scattering events, executed by
the droplet and enabled by its pilot wave. Such looping trajectories
are familiar from walker--pillar scattering~\citep{Harris2018}: here
$B_1$ and $B_2$ play the role of the pillar. The pilot wave at
recrossing (Fig.~\ref{fig:wavefield}c) suggests that the loops rely
on the walker's tendency to turn on arcs of radius $R\sim\lf/2$.

\begin{figure}[tbp]
\centering
\includegraphics[width=\linewidth]{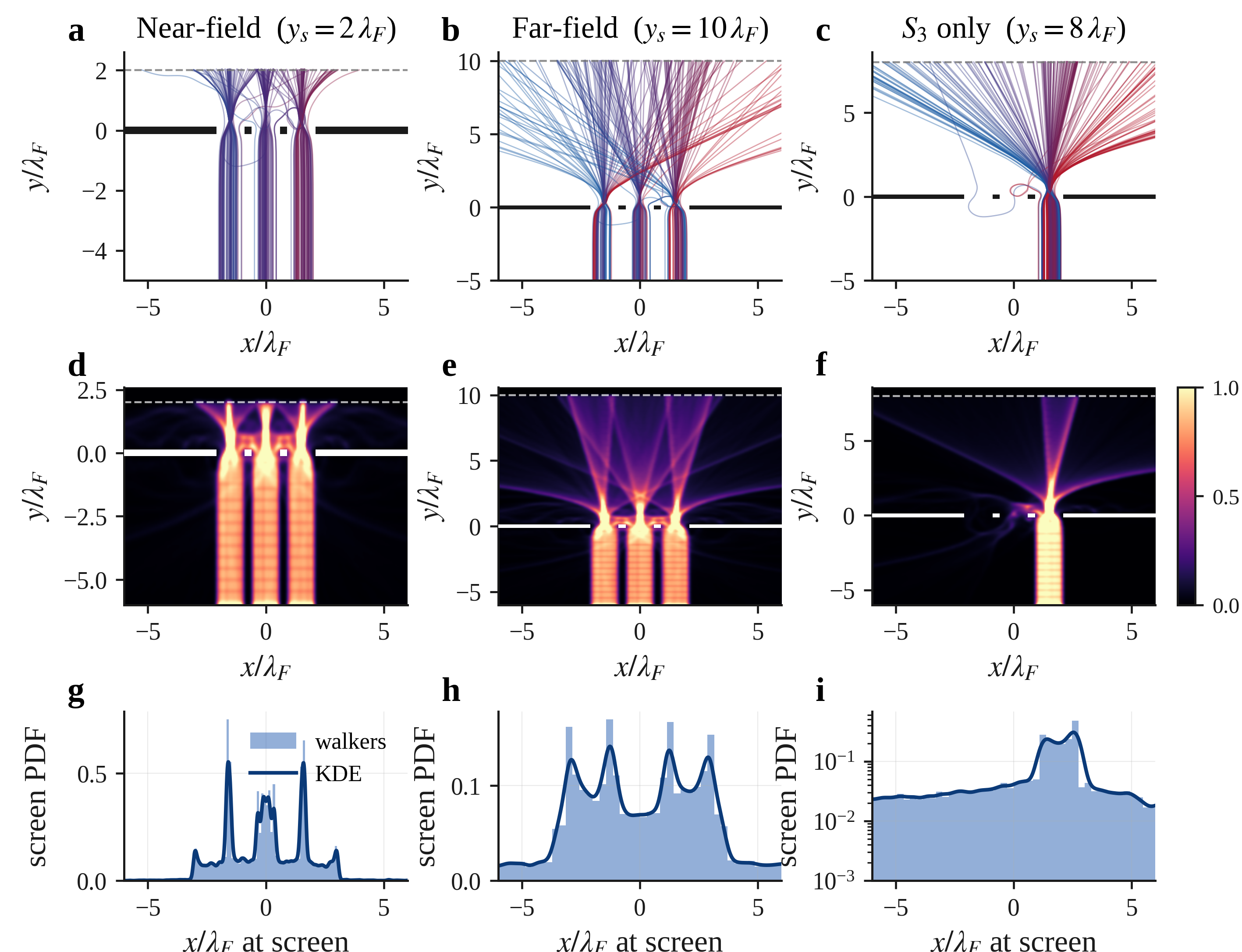}
\caption{\textbf{Triple-slit diffraction: near field, far
field and single-slit launches.}
(a--c)~Walker trajectory bundles; (d--f)~empirical probability
density $p(x,y)$; (g--i)~screen histograms with overlay of kernel-density
estimate (KDE). Columns correspond to the near field
($y_s=2\,\lf$, $N=14{,}890$; panels a,d,g), the far field
($y_s=10\,\lf$, $N=13{,}105$; panels b,e,h), and walkers launched
exclusively through $S_3$ ($y_s=8\,\lf$, $N=10{,}307$; panels
c,f,i). The near field (g) shows three intensity peaks with secondary
maxima between the slits; the far field (h) develops the multi-fringe
diffraction envelope with side lobes extending to
$|x|\approx 5\,\lf$.}
\label{fig:datagrid}
\end{figure}

Figure~\ref{fig:datagrid} demonstrates that our model captures
a form of triple-slit diffraction and single-particle interference. 
The near-field column (panels a,d,g) shows three intensity peaks in the density directly at a distance of $2 \lambda_F$ behind the slit 
centres (d), along with secondary maxima. 
The far-field column (panels b,e,h) reveals the
multi-fringe diffraction envelope (h) on a screen $10 \lambda_F$ behind the slit, with side lobes extending to $|x|\approx 5\,\lf$. 
The central lobe and the secondary side lobes follow the same 
qualitative form as the envelope reported
by Pucci \emph{et al.}~\citep{Pucci2018Faraday} for drops passing
through a single slit, a form now modulated by the three-slit interference.
In the third column (panels c,f,i), we report the fate of walkers 
launched exclusively through $S_3$. The screen probability density (panel i, logarithmic scale) then shows arrivals spread behind all three slits, and the
associated trajectories include a small population of looped
trajectories that recross the barrier plane
(Supplementary Fig.~\ref{fig:allloops}a).
These data demonstrate single-particle interference in the walker 
statistics: the presence of neighbouring apertures reshapes the
distribution $p(x)$.

\begin{figure}[tbp]
\centering
\includegraphics[width=\linewidth]{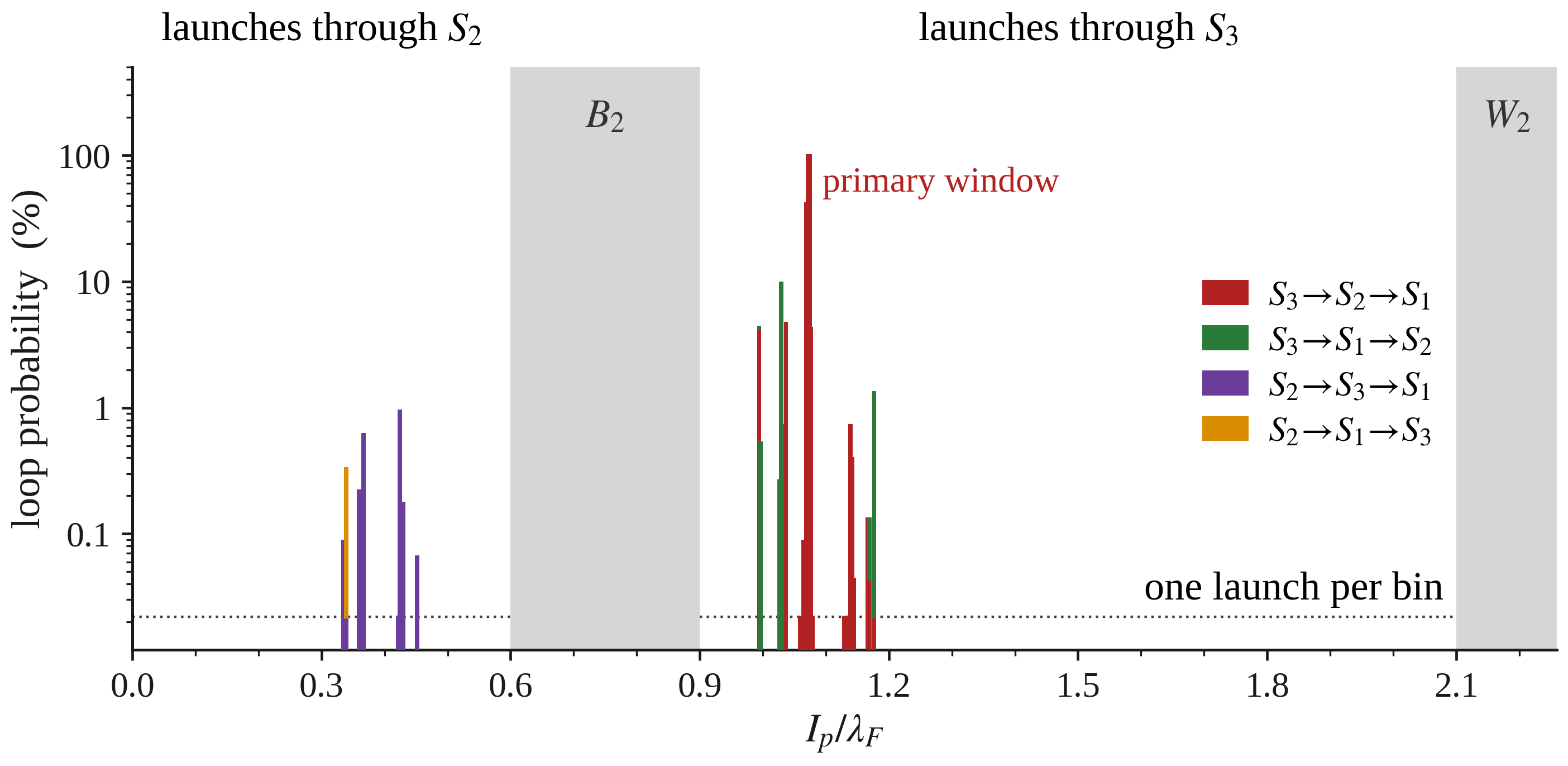}
\caption{\textbf{Dependence of loop probability on impact
parameter.} Loop probability, the percentage of launches that thread
all three slits and reach the screen, as a function of the impact
parameter $I_p$, the perpendicular distance between the initial
trajectory and the centreline of the slit array. We report the
results of launches through the right side of the central slit $S_2$
($0 \le I_p \le 0.6\,\lf$) and through the rightmost slit $S_3$
($0.9\,\lf \le I_p \le 2.1\,\lf$); results for $I_p<0$ are
symmetric. The grey bands mark the barrier segment $B_2$ between the
slits $S_2$ and $S_3$, and the outer wall $W_2$. Launches are
distributed uniformly across each slit, less a margin of
$0.05\,\lf$ at its edges, and are spaced by $5.5\times10^{-7}\,\lf$:
$1{,}000{,}001$ launches across the right side of $S_2$ give rise to
$112$ loops, while $2{,}000{,}001$ across $S_3$ result in $12{,}242$
(Supplementary Secs.~S1 and S2). The launches are sorted into bins of
width $0.0025\,\lf$ in $I_p$, each holding $4{,}545$ or $4{,}546$
launches, and the loop probability is the percentage of the launches
from each bin that loop. Note the logarithmic scale; the dotted line
marks one launch per bin. Bars are colour coded according to
itinerary. Loops arise only within narrow windows of $I_p$; no walker
launched with $I_p<0.33\,\lf$ loops. For $S_3$ launches, every launch
within the primary window $I_p\in[1.069,1.075]\,\lf$ loops;
moreover, this window holds $91\,\%$ of the
loops. The satellite windows, among them those executing
$S_3\!\to\!S_1\!\to\!S_2$ (green), are separated by
$\Delta I_p\simeq0.036\,\lf$. For $S_2$ launches, loops are less
likely, with a maximum probability of $0.95\,\%$.}
\label{fig:floopip}
\end{figure}

We proceed by quantifying the likelihood of the looping trajectories
by evaluating the fraction, $\floop$, of the total that do so.
Only trajectories that arrive at the screen are counted: trajectories that
recross the barrier but do not reach the screen are excluded 
(Methods).
We first consider the fraction of walkers that first pass through $S_3$, 
but subsequently pass through $S_2$ and $S_1$ before reaching
the far-field screen. For the geometry considered 
$f_{\mathrm{loop},3} = 107/13{,}334 = 0.80\,\%$ (Methods).
Figure~\ref{fig:floopip} makes clear that the likelihood of looping 
depends strongly on the impact parameter: indeed, looping trajectories 
arise exclusively from within narrow windows of $I_p$. Specifically, 
every launch from within the primary window ($I_p\in[1.069,\,1.075]\,\lf$, red) 
results in a looped trajectory.
Figure~\ref{fig:floopip} is constructed from uniform grids of launches,
spaced $5.5\times10^{-7}\,\lf$ apart in $I_p$ and sorted into bins of
width $0.0025\,\lf$, each of which holds approximately $4{,}500$
launches. The ordinate, the loop probability, is the percentage of the
launches in each bin that loop, plotted on a logarithmic scale. 
All $11{,}171$ launches of the grid that fall within the
primary window loop, without exception, so that the two bins lying
wholly within it register $100\,\%$.
The data are colour coded according to the form of the loops, which 
makes clear that the most prevalent form of loop is the $S_3\!\to\!S_2\!\to\!S_1$
form illustrated in Fig.~\ref{fig:diffraction}b,c, the majority of which 
come from the primary window.

A series of narrow windows contribute relatively few loops, among
them those of the form $S_3\!\to\!S_1\!\to\!S_2$ (green).
Of the $107$ loops of the $2\times10^{4}$ ensemble launched through $S_3$, $101$ execute a
similar double-spiral path, $S_3\!\to\!S_2\!\to\!S_1$. The
remaining six, all launched from the satellite windows, thread the
apertures in the reverse order, $S_3\!\to\!S_1\!\to\!S_2$ (green in
Fig.~\ref{fig:diffraction}c and Supplementary Fig.~\ref{fig:allloops}a),
while two of them execute additional spiral turns. 
The probability of a loop trajectory for particles passing through $S_3$
is thus $107/20{,}000 = 0.54\,\%$ (Methods). This figure counts the loops among all
launched walkers, whereas $f_{\mathrm{loop},3}$ counts them among the walkers that reach the screen.
Both are used in the comparison with quantum backflow (Supplementary Sec.~S5),
because the backflow probability is likewise expressed as a fraction of the
entire incident wavepacket, not only of the component that is transmitted.

Walkers launched through the central aperture $S_2$
(Fig.~\ref{fig:diffraction}d and Supplementary Sec.~S2) may undergo the same
spiral scattering, with a slightly different outcome. About $15\,\%$ of them
spiral around the barrier segment nearest their launch point and return
through the adjacent outer slit, after which none reaches the screen.
Trajectories that thread all three apertures persist, among them
$S_2\!\to\!S_3\!\to\!S_1$ and
$S_2\!\to\!S_1\!\to\!S_3$, account for only $0.018\,\%$ of particles
arriving at the screen. For comparison, on a grid of launches of the same 
spacing, loops account for $0.92\,\%$ of the walkers 
launched through $S_3$ that reach the screen, consistent with the value
$f_{\mathrm{loop},3}=0.80\,\%$ ($95\,\%$ confidence interval (CI) $[0.65,0.96]\,\%$) obtained above from the smaller ensemble. The  
windows of $I_p$ prone to looping
(Fig.~\ref{fig:floopip}, left) lie 
within $0.3\,\lf$ of the barrier about which the walker first spirals.
Nevertheless, these windows are narrower and hold far fewer loops than those of the $S_3$ launches. 

\begin{figure}[tbp]
\centering
\includegraphics[width=\linewidth]{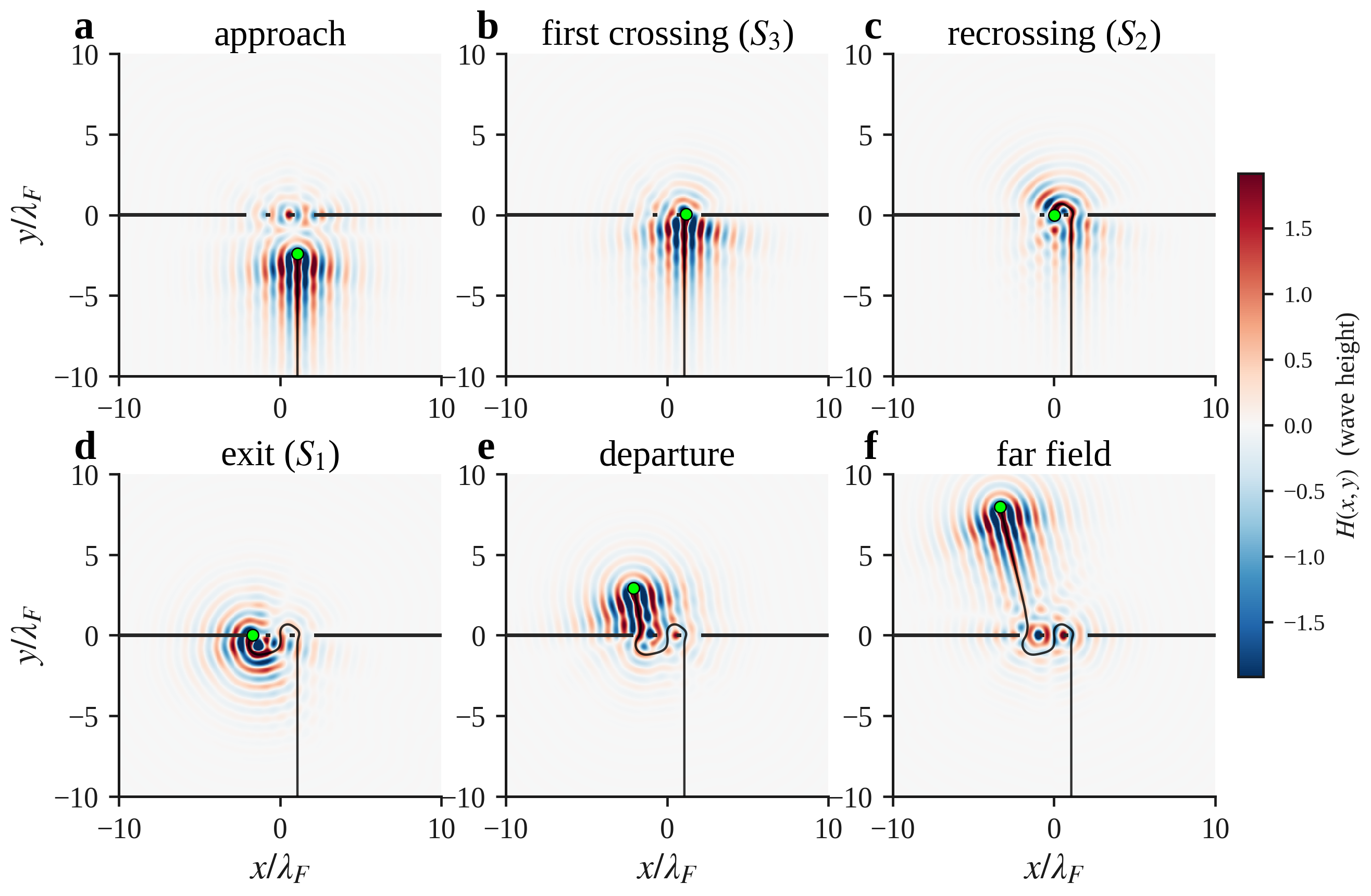}
\caption{\textbf{The pilot wave of a looping walker.}
Reconstructed wave height $H(x,y)$ (red positive, blue negative) of one
looping walker at six stages of its transit, with the trajectory
(black) and the current droplet position (green) overlaid. The field is
the sum of the waves generated by the droplet and barrier (Methods).
(a)~The walker approaches the barrier near $S_3$, dressed in its pilot
wave.
(b)~First crossing (of $S_3$): the pilot wave is distorted by the 
slit arrangement.
(c)~The droplet spirals about $B_2$ and recrosses $S_2$.
(d)~A second spiral about $B_1$ directs it through $S_1$.
(e)~The walker departs the barrier region.
(f)~The walker proceeds towards the screen, its pilot wave
relaxing to that of a free walker.}
\label{fig:wavefield}
\end{figure}

Direct walkers (a single forward crossing) and loopers ($\geq 2$
crossings) encounter markedly different wave fields at their first
arrival at the barrier plane. The structure
encountered by the direct population is relatively flat, though
characterized by the corrugation responsible for the channelling
reported by Pucci \emph{et al.}~\citep{Pucci2025}.
Conversely, when a looper first arrives, the waves are relatively large 
in amplitude, and so serve to deflect the drop onto a spiral about $B_2$.
A quantitative reconstruction of the barrier wave field for direct crossings and
loopers is given in Supplementary Fig.~\ref{fig:vprofiles}.

\section*{Discussion}

The central result of this paper is the discovery that, in pilot-wave
hydrodynamics, rare but robust looped trajectories may arise in triple-slit
diffraction. With the geometry considered, $0.80\,\%$ of the walkers that 
reach the screen thread all three apertures in turn, executing the spiral-scattering dynamics
familiar from walker--pillar interactions~\citep{Harris2018}.
Indeed, the slit geometry was chosen with the spiral scattering in
mind. The loops arise only from a number of discrete windows of
$I_p$ (Fig.~\ref{fig:floopip}), the bulk of them from the primary windows of the 
outermost slits. Walkers launched through the central slit loop far more rarely 
(Figs.~\ref{fig:diffraction}d and \ref{fig:floopip}): their first spiral
returns them through an outer slit, thus precluding a second spiral scattering event. Nevertheless, some particles are able to recross the plane of the slit via a relatively unlikely path.

The mechanism is geometric: the instantaneous wave field
$\hwave(\boldsymbol{x},t)$ is constrained to vanish on the
barrier~\citep{Pucci2025}, so the
barrier-source contribution $H_s(\boldsymbol{x},t)$ develops a spatial
structure dictated by the slit geometry. Spiral scattering is
therefore a geometry-specific feature of pilot-wave dynamics.
One-parameter sweeps about the reference geometry (Supplementary
Sec.~S3 and Supplementary Fig.~\ref{fig:landscapeSI}) show that the
loop fraction varies strongly with slit separation, slit width and
memory. Among the geometries examined, the chosen one lies close to a
local maximum of the loop fraction with respect to both the slit
separation and the slit width. The loop fraction also depends on the
kernel decay length: loops persist provided the kernel is sufficiently
long to span adjacent slits ($\delta\gtrsim1.5\,\lf$; Supplementary
Sec.~S4 and Supplementary Table~\ref{tab:delta}).

It is interesting to compare the probabilities of our loop trajectories with those arising in an isometric quantum setting. This isometry is achieved by identifying $\lambda_{\mathrm{dB}}$ with $\lf$, as is customary in hydrodynamic quantum analogues \citep{Bush2015AnnRev,BushOza2021}.
We note that the comparison is qualitative. 
In the hydrodynamic system, the probability is deduced from the trajectories,
yielding $f_{\mathrm{loop},3}=0.80\,\%$. In quantum mechanics, it is assessed through 
an evaluation of the integrated backward flux within a broadened Schr\"{o}dinger
wavepacket, yielding $\Pback^{+}=0.86\,\%$ (see Supplementary Sec.~S5).
Nevertheless, the fraction of loop trajectories is comparable. What the comparison
establishes is a regime in which loop trajectories are comparably
likely in the two systems. Most importantly, exotic loop
trajectories, predicted but not directly observable in quantum
diffraction, arise naturally and are directly observable in
pilot-wave hydrodynamics.

While we attempted the corresponding experiment in the laboratory,
difficulties arose.
First, the looping spirals span several Faraday wavelengths, requiring
a bath substantially larger than those currently 
available~\citep{Bush2015AnnRev,Pucci2018Faraday}. Second, the loops
are rare: with the chosen geometry, approximately one launch in 187
loops and terminates on the screen (representing 1/125 of the walkers
that reach the screen). Thus, resolving the looping probability with confidence
would require on the order of $10^{4}$ experimental launches. These constraints served to motivate the numerical approach taken here. Nevertheless, we did observe some vestiges of the 
loops in the laboratory, more robust versions of which could be captured on a larger bath.

In the photonic three-slit
experiment of Maga\~{n}a-Loaiza \emph{et al.}~\citep{MaganaLoaiza2016},
the loops are inferred indirectly from far-field statistics, 
and are enhanced via engineered plasmonic near fields.
In our study, the loops are
tracked directly in real space, and arise naturally in response to the droplet's pilot wave. Whether a deeper structural identity links the
walker's spiral-scattering dynamics to the quantum states that
maximize backflow, or whether the correspondence is specific to
the three-slit geometry, is left as a question for future work.

\section*{Methods}

\paragraph{Stroboscopic model.}
We integrate the stroboscopic model in the canonical formulation of
Pucci, Bellaigue, Cirimele, Al\`{i} \& Oza~\cite{Pucci2025}, i.e.\
Eqs.~\eqref{eq:eom}--\eqref{eq:hwave} of the main text supplemented by
the instantaneous Dirichlet condition $\hwave(\boldsymbol{x}_j,s)=0$
on each barrier source $\boldsymbol{x}_j$. The dimensionless memory
parameter $\Me \equiv \TM/\TF = T_d/[\TF(1 - \gamma/\gamma_F)]$, with
$T_d$ the bath wave decay time in the absence of forcing and
$\gamma/\gamma_F$ the dimensionless vibrational forcing. All lengths are
non-dimensionalized by the Faraday wavelength $\lf$, all times by the
bouncing period $\TF$. Eq.~\eqref{eq:eom} is advanced in time by a
damped-Verlet update, the standard discretization of
Eq.~\eqref{eq:eom} with linear drag:
$\Xp^{n+1} = c_1 \Xp^{n} - c_2 \Xp^{n-1} - c_3\,\nabla\Hfield(\Xp^{n}, n\TF)$
with $c_1 = (2\kappa+1)/(\kappa+1)$, $c_2 = \kappa/(\kappa+1)$,
$c_3 = \alpha/(\kappa+1)$ and dimensionless drag $\kappa = m/(D\TF) = 0.2$. The single-impact wave kernel is
$\hzero(r)=J_0(2\pi r)\,e^{-r/\delta}$, with the spatial decay length
$\delta$ controlling the range of the boundary-induced field (the
symbol $d$ being reserved for the slit separation throughout). In the
runs reported here we take $\delta=1.5\,\lf$, corresponding to a
moderately strongly damped Faraday wave. The sensitivity to $\delta$
is quantified directly under the headline protocol (Supplementary
Table~\ref{tab:delta}): the loop fraction collapses for
$\delta\leq1.2\,\lf$ ($0.06$--$0.07\,\%$), consistent with the
mechanistic requirement that the kernel span adjacent apertures
($\delta\gtrsim d$), and persists with growing amplitude at
$\delta=2.0\,\lf$ ($0.98\,\%$, CI $[0.71,1.28]\,\%$) and
$\delta=2.5\,\lf$ ($3.2\,\%$, CI $[2.0,4.4]\,\%$). In the
$\delta\approx3\,\lf$ window of the diffraction runs of
Ref.~\cite{Pucci2025} ($\gamma/\gamma_F=0.95$), annihilation at the
barrier dominates ($96\,\%$ of launches) and few walkers reach the far
screen, so that window requires a protocol adapted to the heavy
attrition to be quantified reliably. Other parameters:
walking speed $V_0=0.05\,\lf\,\mathrm{step}^{-1}$, memory truncation
$n_m=5\,\Me$. The dimensionless wave-force amplitude $\alpha$ in the
stroboscopic Verlet scheme is fixed self-consistently by requiring
the unforced walker to drift at $V_0$:
$\alpha = -V_0 / \sum_{k=1}^{n_m} \hzero'(V_0 k)\,e^{-k/\Me}$.
Barrier geometry: slit width $L=1.2\,\lf$ and barrier-segment width
$b=0.3\,\lf$, so that the centre-to-centre slit separation is
$d=L+b=1.5\,\lf$.

\paragraph{Boundary treatment.}
Following Pucci \emph{et al.}~\cite{Pucci2025}, the impenetrable barrier
at $y=0$ is parametrized by $N_s=64$ point sources. Each outer wall
carries $30$ sources at spacing $\Delta x_s=\lf/2$, the first at the
edge of the adjacent slit ($x=\pm2.1,\,\pm2.6,\,\ldots,\,\pm16.6\,\lf$).
Each interior segment, whose width $b=0.3\,\lf$ is less than
$\Delta x_s$, carries two sources, one at each of its edges
($x=\pm0.6$ and $\pm0.9\,\lf$). At every stroboscopic step the source weights
$\vec W(t)$ are updated via
\[
\vec W(t)
  \;\leftarrow\;
  e^{-1/\Me}\,\vec W(t-\TF)
  \;+\;
  \vec M^{-1}\!\bigl(-\hzero(\vec r_{\mathrm{walker}\to\mathrm{src}})\bigr),
\qquad
M_{ij}=\hzero(|\boldsymbol{x}_i-\boldsymbol{x}_j|),
\]
which enforces the instantaneous Dirichlet condition
$\hwave(\boldsymbol{x}_j,t)=0$ at every source $\boldsymbol{x}_j$
and propagates each impact's contribution to the source amplitudes
with the same memory kernel as the walker's own wave. The time-averaged source field
$H_s(\boldsymbol{x},t) = \sum_j W_j(t)\,\hzero(|\boldsymbol{x}-\boldsymbol{x}_j|)$
is therefore obtained at no additional cost from the propagated
weights. We verified by direct
numerical evaluation that the residual $\max_j|\hwave(\boldsymbol{x}_j,t)|$
is suppressed by more than four orders of magnitude relative to the
single-impact amplitude over the full integration window.

\paragraph{Crossing detection and classification.}
At each step we test whether the segment $(\vec x_n,\vec x_{n+1})$
crosses $y=0$. A crossing inside a slit aperture is admitted;
otherwise the walker is annihilated. Walkers are classified as
\emph{direct} (exactly one forward crossing through the entry slit,
$S_3$ unless stated otherwise) or \emph{looped} (two or more
crossings, with at least one through another slit). A full loop
requires at least one crossing recorded at each of $S_1$, $S_2$ and
$S_3$; for launches through $S_3$, in the order $S_3\to S_2\to S_1$
or $S_3\to S_1\to S_2$ (the itineraries of launches through $S_2$
are catalogued in Supplementary Sec.~S2).

\paragraph{Loop-fraction measurement.}
$N=2\times10^{4}$ walkers are launched uniformly from $S_3$ and tracked
until each reaches the screen at $y_s=10\,\lf$, leaves the
integration domain (laterally beyond $|x|=20\,\lf$, or swept back
more than $5\,\lf$ behind the launch line), or is annihilated
against the barrier. Each launch is
set by its \emph{impact parameter} $I_p$, the launch abscissa
measured from the centreline of the slit array and taken positive
towards $S_3$; launches through
$S_3$, whose centreline lies at $I_p=d=1.5\,\lf$, sample $I_p$
uniformly over $[0.95,2.05]\,\lf$,
the aperture less $0.05\,\lf$ at either edge, and whether a walker
ultimately loops depends on $I_p$. Of the walkers
that reach the screen ($13{,}334$ of the $N=2\times10^{4}$
launches), $f_{\mathrm{loop},3}$ is the fraction whose recorded
trajectory crosses $S_3$ and subsequently both $S_2$ and $S_1$, in
either order; loopers that
recross the barrier but are annihilated or escape before arriving are
not detected and are excluded. The $95\,\%$ confidence interval is a
bootstrap over $B=2{,}000$ resamples of the screen-arrival list,
giving $f_{\mathrm{loop},3}=0.80\,\%$, CI $[0.65,0.96]\,\%$. The
$I_p$-resolved statistics of Supplementary Sec.~S1 use an independent
ensemble of $N=10^{6}$ launches under the identical protocol and a
uniform grid of $2{,}000{,}001$ launches; in the former, all $5{,}652$ launches falling
within the window $I_p\in[1.069,\,1.075]\,\lf$ loop without
exception. The model is symmetric under reflection about the
centreline of the slit array, which exchanges $S_1$ and $S_3$: the
outer-slit ensembles were computed for launches to the left of the
centreline and are reported throughout as their mirror images, as
launches through $S_3$. On the grid of launches through $S_2$
(Supplementary Sec.~S2) the mirror symmetry was verified to hold
exactly for all $10^{6}$ pairs of launches at $\pm I_p$.
Integration uses a Numba-JIT walker.

\paragraph{Walking speed near the barrier.} Across the representative
direct and looped ensembles, the per-bounce walking speed within
$|y|<\lf$ of the barrier remains within approximately $20\,\%$ of the
free walking speed $V_0$ (5th--95th percentile; direct walkers
$-19\,\%$ to $+5\,\%$, loopers $-20\,\%$ to $+13\,\%$), so the
walker's wavelength remains well defined throughout the interaction.

\paragraph{Mechanism sample (Supplementary Fig.~\ref{fig:vprofiles}).}
Representative direct ($n=200$) and looped ($n=31$) walkers are drawn
from the full ensemble at the chosen geometry. The transient self-potential
$V(x;t_*^{(1)}) \equiv H_s\bigl((x,0),\,t_*^{(1)}\bigr)$ --- the $y=0$
cut of the time-averaged barrier-source field at the walker's first
arrival at the barrier plane --- is reconstructed at the first
crossing event of each walker. Ensemble averages and
$\pm 1\sigma$ bands are computed over these populations. The looper
sample is small ($n=31$).

\bibliographystyle{unsrtnat}
\bibliography{ARFMBib,npref}

\section*{Acknowledgements}
\noindent
The authors thank A.~U.~Oza and G.~Pucci for valuable discussions, 
and Cl\'{e}mence Fl\'{e}chelle for her attempts to capture the loops 
experimentally. J.W.M.B. gratefully acknowledges the generous financial 
support of the National Science Foundation through grant CMMI-2609598 and 
the Office of Naval Research through grant N000014-24-1-2232. 
G.G.R.\ gratefully acknowledges the
support of a Sloan Fellowship awarded through MIT.

\section*{Author contributions}
\noindent
G.G.R.\ designed the simulations, wrote the numerical codes
(stroboscopic walker, boundary-source memory representation,
Schr\"{o}dinger split-step integrator)
and produced the figures. J.W.M.B.\ supervised the project, framed
the connection to quantum backflow and the photonic three-slit
literature, and contributed to the physical interpretation. Both
authors wrote the manuscript.

\section*{Competing interests}
\noindent
The authors declare no competing interests.

\section*{Data availability}
\noindent
The trajectories of all $107$ looped walkers, the loop-fraction and
impact-parameter ensembles for launches through $S_2$ and $S_3$, the parameter-sweep logs, and
the analysis code supporting this study will be deposited in a public
repository (DOI to be inserted upon acceptance); until then they are
available from the corresponding author on request.

\setcounter{figure}{0}
\setcounter{table}{0}
\setcounter{equation}{0}
\renewcommand{\thefigure}{S\arabic{figure}}
\renewcommand{\thetable}{S\arabic{table}}
\renewcommand{\theequation}{S\arabic{equation}}
\renewcommand{\theHfigure}{S\arabic{figure}}
\renewcommand{\theHtable}{S\arabic{table}}
\renewcommand{\theHequation}{S\arabic{equation}}
\section*{Supplementary Information}

\noindent
This supplement details the loop statistics (Secs.~S1 and S2), their
dependence on the model parameters (Secs.~S3 and S4) and the
comparison with quantum backflow (Sec.~S5). Sections S1, S3 and S4,
and the walker statistics used in Sec.~S5, concern walkers launched
through the outer slit $S_3$ or, equivalently, $S_1$; launches through
the central slit are treated in Sec.~S2. The emphasis reflects their
relative weight: on grids of launches of equal spacing (Secs.~S1 and
S2), the two outer slits give rise to $24{,}484$ three-slit loops and
the central slit to $224$, so that $99\,\%$ of the loops originate at
the outer slits.

\subsection*{S1. Loop probability versus impact parameter}

\begin{figure}[htbp]
\centering
\includegraphics[width=\linewidth]{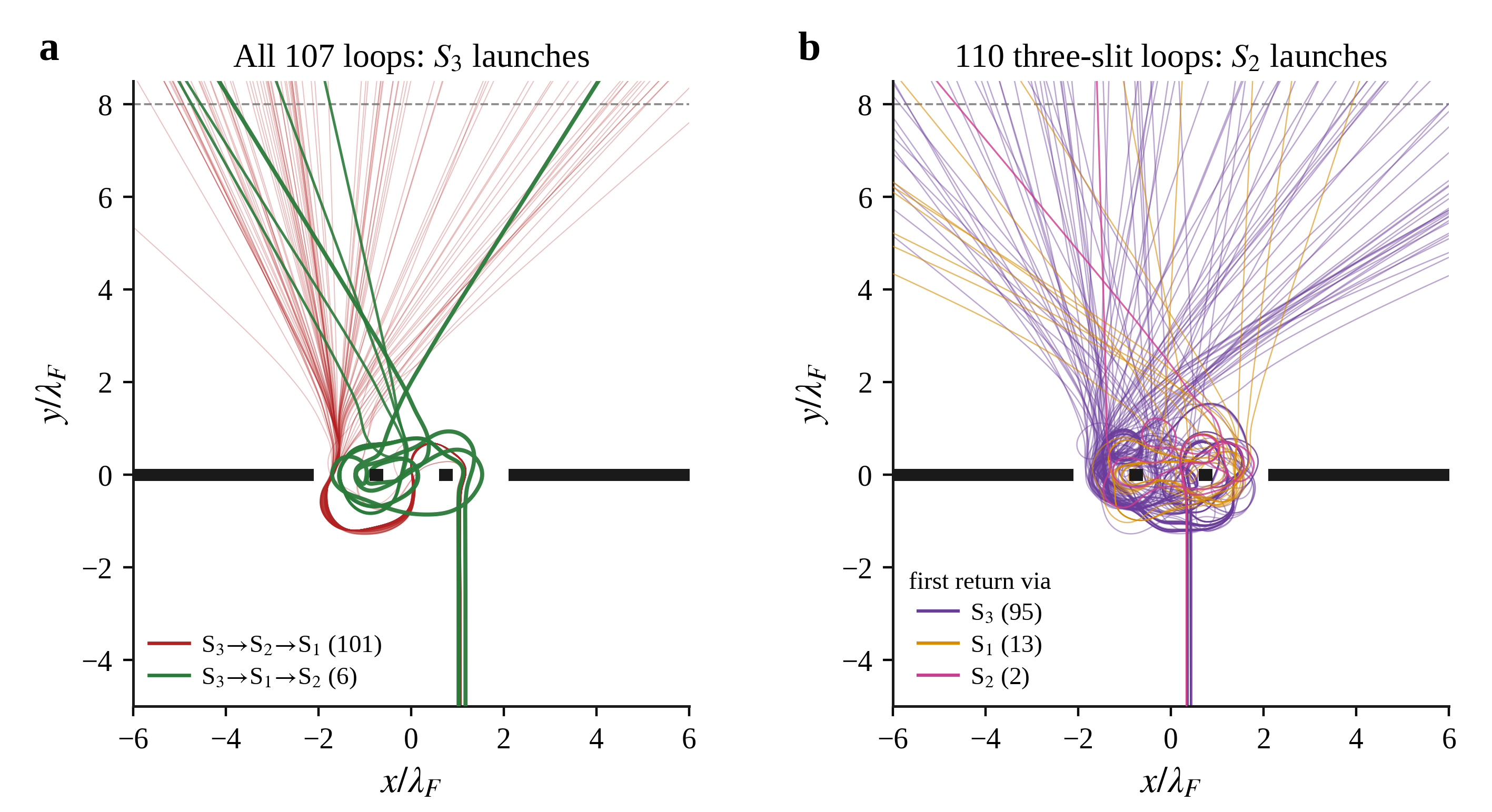}
\caption{\textbf{Three-slit loops of the two ensembles.}
(a)~The $107$ loops of the $2\times10^{4}$-launch ensemble through
$S_3$ (Fig.~\ref{fig:diffraction}c of the main text), coloured by
itinerary: $S_3\!\to\!S_2\!\to\!S_1$ (red, $101$) and
$S_3\!\to\!S_1\!\to\!S_2$ (green, 6, the two darker ones execute
additional spiral turns).
(b)~Launches through the right half of $S_2$ that thread all three
slits before reaching the screen: $110$ of the $112$ of the grid of
Sec.~S2,
coloured by the slit of first return: the near outer slit
$S_3$ (purple, $95$), the far outer slit $S_1$ (orange, $13$) or $S_2$
itself (magenta, $2$). The dashed line marks $y=8\,\lf$.}
\label{fig:allloops}
\end{figure}

\noindent
Two ensembles of launches through $S_3$ underlie the $I_p$-resolved
statistics of the main text. The first comprises $N=10^{6}$ launches
distributed uniformly in $I_p$ under the protocol of Methods; it yields
$6{,}220$ loops among $666{,}437$ screen arrivals, a loop fraction of
$0.93\,\%$ ($95\,\%$ confidence interval, CI, $[0.91,0.96]\,\%$), at
the upper end of the interval obtained from the $2\times10^{4}$
ensemble of the main text. The second is the uniform grid of
$2{,}000{,}001$ launches on which Fig.~\ref{fig:floopip} of the main
text is based; it yields $12{,}242$ loops among $1{,}333{,}897$ screen
arrivals ($0.918\,\%$, CI $[0.902,0.934]\,\%$). The two ensembles
yield the same windows. The primary window spans
$I_p\in[1.0690,\,1.0751]\,\lf$, a width of $0.0061\,\lf$. Satellite
windows lie at $I_p\approx0.994\,\lf$ and at $1.027$--$1.037\,\lf$,
with weaker ones between $1.138$ and $1.176\,\lf$; none holds more
than $6\,\%$ of the loops. The itinerary is set by the window: the
primary window produces $S_3\!\to\!S_2\!\to\!S_1$ exclusively, whereas
the six loops of the $2\times10^{4}$ ensemble that thread the slits in
the reverse order (Supplementary Fig.~\ref{fig:allloops}a) originate
in satellite windows, five near $I_p\approx1.028\,\lf$ and one near
$1.176\,\lf$. The wave field met at the first crossing of the barrier
by direct walkers and by future loopers is compared in Supplementary
Fig.~\ref{fig:vprofiles}.

In all ensembles the launches are distributed uniformly across the
slit, less a margin of $0.05\,\lf$ at each of its edges, where wave
sources are located (Sec.~S3). The margin excludes only launches that
do not reach the screen: of $4{,}000$ launches within each margin
(those of $S_3$ adjacent to $B_2$ and to $W_2$, and that of $S_2$
adjacent to $B_2$), every walker passed through the slit and was then
annihilated at the barrier. The loop fraction among screen arrivals is
therefore unaffected by the margin, while the loop probability per
launch would be reduced by a factor $11/12$ were the margins included.

\begin{figure}[htbp]
\centering
\includegraphics[width=0.62\linewidth]{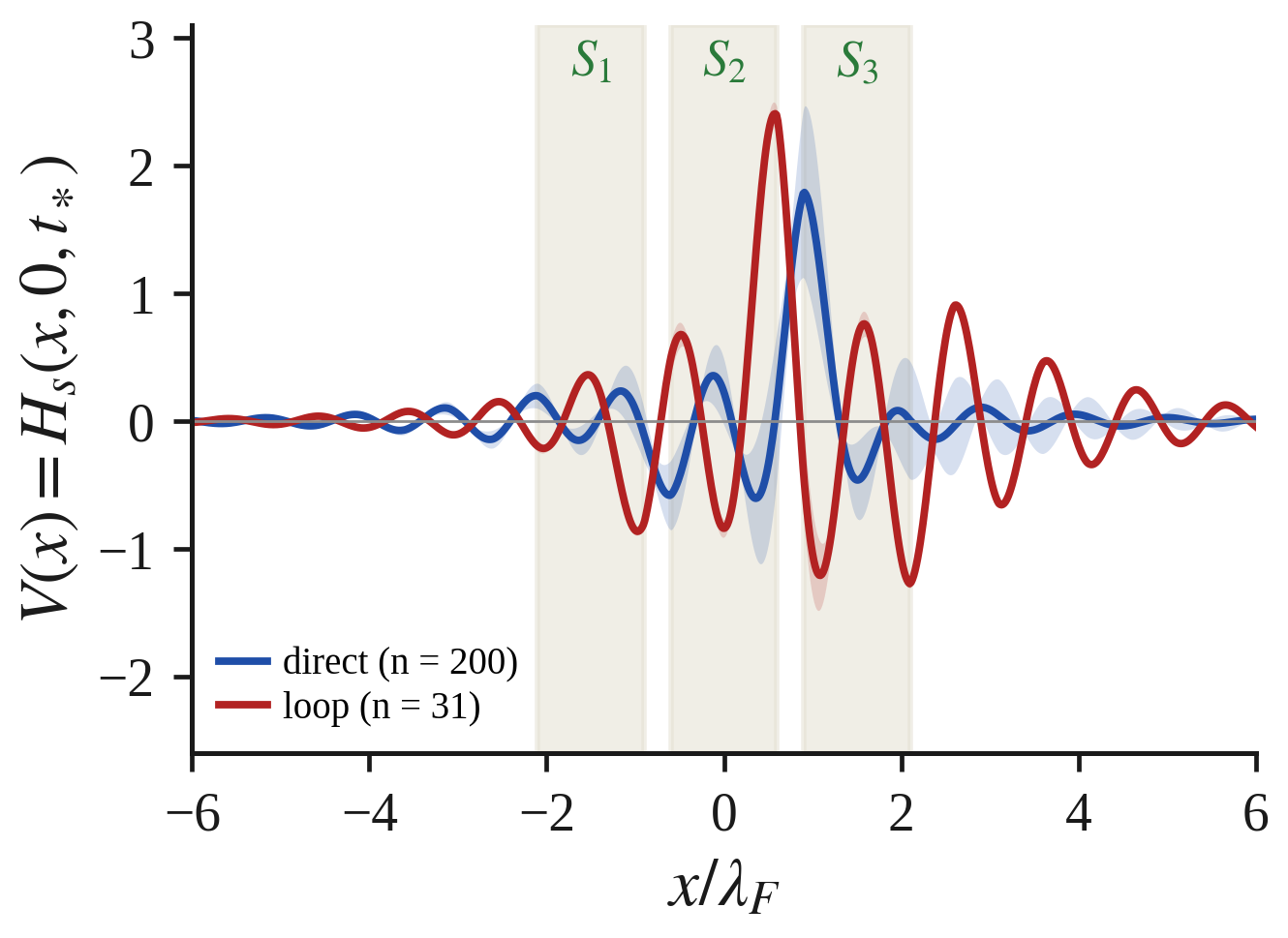}
\caption{\textbf{Ensemble-averaged transient self-potential.}
$V(x;t_*^{(1)})$, the barrier-source wave field across the barrier at
the first crossing (Methods), for the direct (blue, $n=200$) and
looped (red, $n=31$) populations of walkers launched through $S_3$;
shaded bands, $\pm1\sigma$; vertical bands mark the three slits
$S_1$, $S_2$ and $S_3$. Loopers arrive at a pre-existing attractive
structure near $S_2$; the field met by the direct ensemble is weaker
and is concentrated at the edge of $S_3$ nearest $B_2$. $V$ is expressed in units of the single-impact wave
amplitude. The gradient of $V(x)$ on the inner side of $S_3$ deflects
the droplet onto a spiral path.
}
\label{fig:vprofiles}
\end{figure}

\subsection*{S2. Launches through the central slit}

\noindent
Launches through the central slit $S_2$
(Fig.~\ref{fig:diffraction}d and Fig.~\ref{fig:floopip} of the main
text; Supplementary Fig.~\ref{fig:allloops}b) were examined on a
uniform grid of $2{,}000{,}001$ launch positions spanning
$|I_p|\leq0.55\,\lf$ (spacing $5.5\times10^{-7}\,\lf$), all else being
as in Methods. As trajectories launched at $\pm I_p$ are mirror images
of one another (Methods), we report the right half of the grid,
$0\leq I_p\leq0.55\,\lf$ ($1{,}000{,}001$ launches), for which $S_3$
is the \emph{near} outer slit and $S_1$ the \emph{far} one. Confidence
intervals are $95\,\%$ Wilson intervals.

Of these launches, $62.0\,\%$ reach the screen, $28.9\,\%$ are
annihilated at the barrier and $9.1\,\%$ leave the integration domain.
Every walker launched with $I_p<0.326\,\lf$ crosses the barrier once
and reaches the screen; walkers recross only if launched with
$I_p\geq0.333\,\lf$, that is, within $0.27\,\lf$ of the edge of the
slit. Those that recross ($18.9\,\%$ of the launches) do so
predominantly through the near outer slit and are then almost always
lost: $15.28\,\%$ (CI $[15.21,15.35]\,\%$) of the launches follow
$S_2\!\to\!S_3$ and cross no more. Of the walkers whose first return
is through the adjacent slit, $0.7\,\%$ reach the screen, against
$7.3\,\%$ for launches through $S_3$.

Two itineraries of three crossings thread all three slits:
$S_2\!\to\!S_3\!\to\!S_1$ ($217$ launches, of which $46$ reach the
screen) and $S_2\!\to\!S_1\!\to\!S_3$ ($34$ launches, of which $10$
reach the screen). Including longer sequences, $112$ launches thread
all three slits and reach the screen, $0.0181\,\%$ (CI
$[0.0150,0.0217]\,\%$) of the screen arrivals. Itineraries that
recross without visiting all three slits are more common:
$S_2\!\to\!S_3\!\to\!S_2$ ($620$ screen arrivals among $3{,}228$
launches), $S_2\!\to\!S_2\!\to\!S_3$ ($491$ among $10{,}962$) and
$S_2\!\to\!S_3\!\to\!S_3$ ($428$ among $11{,}758$). In all, the
walkers that recross and reach the screen amount to $0.17\,\%$ of the
launches. No itinerary possesses a window comparable to the primary
window of $S_3$: the longest run of consecutive three-slit loops on
the grid spans $6\times10^{-6}\,\lf$, against $6.1\times10^{-3}\,\lf$
for $S_3$.

\subsection*{S3. Barrier discretization and dependence on slit geometry and memory}

\noindent
The barrier is represented by $64$ point sources on the line $y=0$
(Methods). Each outer wall carries $30$ sources spaced $\lf/2$ apart,
the first at the edge of the adjacent slit
($x=\pm2.1,\,\pm2.6,\,\ldots,\,\pm16.6\,\lf$). Each interior segment,
whose width $b=0.3\,\lf$ is less than this spacing, carries two
sources, one at each of its edges ($x=\pm0.6$ and $\pm0.9\,\lf$;
Fig.~\ref{fig:diffraction}a of the main text). The loop fraction is
sensitive to this discretization and to the memory: the values
reported in this study are specific to this arrangement of sources and
to $\Me=50$.

\begin{figure}[htbp]
\centering
\includegraphics[width=0.99\linewidth]{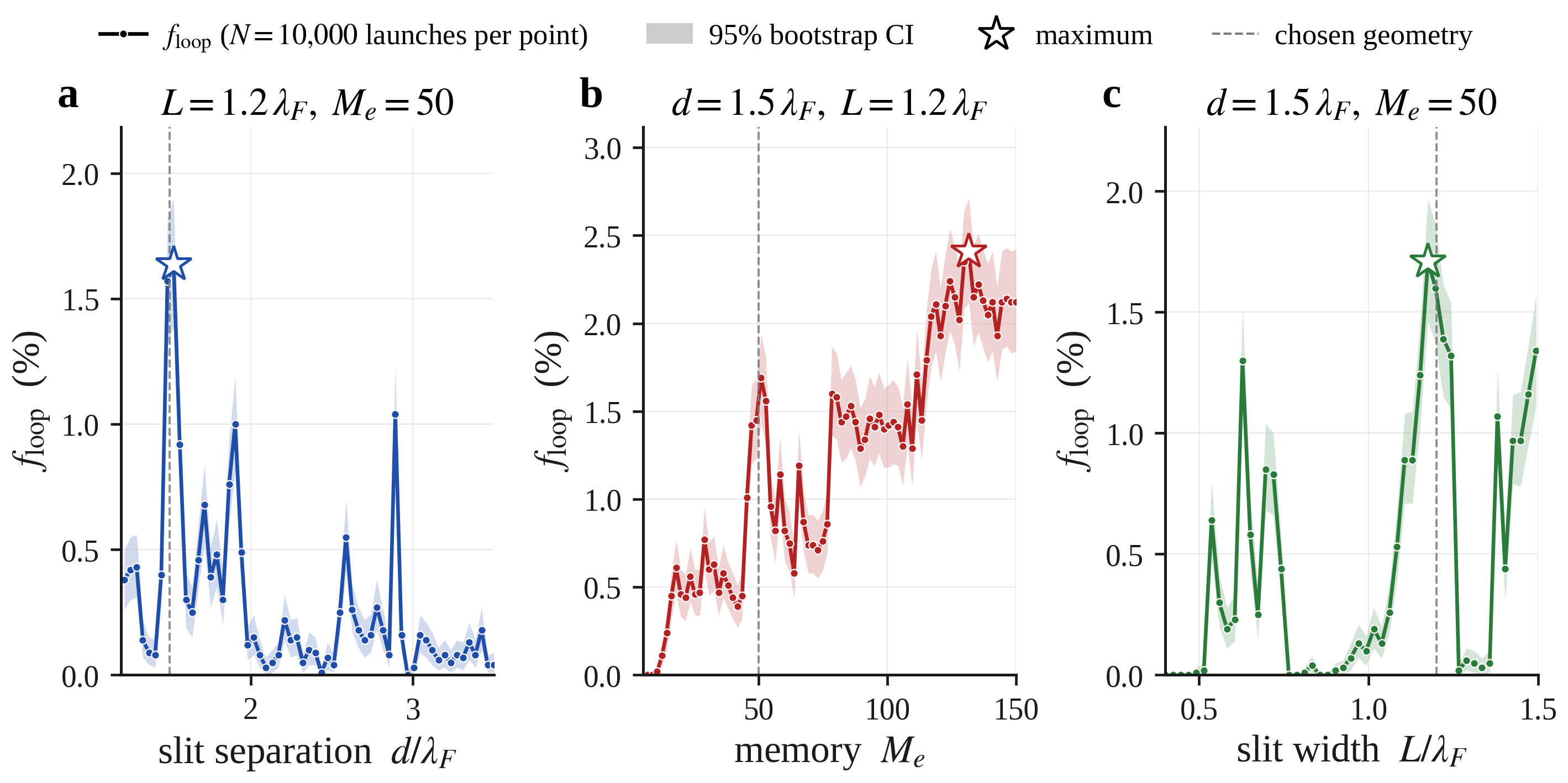}
\caption{\textbf{Dependence of the loop fraction on slit separation,
memory and slit width.} Fraction of launched walkers that thread all
three slits ($N=10{,}000$ launches through $S_3$ per point; shaded
bands, $95\,\%$ bootstrap confidence intervals; detection screen at
$y_s=8\,\lf$; see Sec.~S3 for the relation to the main-text
definition).
(a)~Dependence on the slit separation $d$ at $L=1.2\,\lf$, $\Me=50$,
for $d\geq L$.
(b)~Dependence on the memory $\Me$ at $d=1.5\,\lf$, $L=1.2\,\lf$.
(c)~Dependence on the slit width $L$ at $d=1.5\,\lf$, $\Me=50$, for
$L\leq d$.
Stars mark the maxima; the dashed vertical lines indicate the chosen
geometry, $d=1.5\,\lf$, $\Me=50$ and $L=1.2\,\lf$.}
\label{fig:landscapeSI}
\end{figure}

The dependence on the slit separation, the memory and the slit width
was examined through three one-parameter sweeps about the reference
geometry (Supplementary Fig.~\ref{fig:landscapeSI}; $N=10{,}000$
walkers launched through $S_3$ per value, $95\,\%$ bootstrap
confidence intervals). The sweeps count every three-slit loop among
\emph{launched} walkers, with the screen at $y_s=8\,\lf$, which gives
between $1.45\,\%$ (CI $[1.23,1.68]\,\%$) and $1.69\,\%$ (CI
$[1.44,1.94]\,\%$) at the grid points nearest the chosen geometry. The
main-text value, $0.80\,\%$, counts only the loops that reach the
screen, among screen arrivals at $y_s=10\,\lf$. The difference
reflects the loops that end in annihilation at the barrier or in
escape from the domain: in the ensemble of the main text, $324$ of the
$2\times10^{4}$ launches ($1.62\,\%$) thread all three slits, of which
$107$ reach the screen. Only geometries in which the three slits are
distinct, $d\geq L$, are considered.

The loop fraction is a strongly non-monotonic function of the slit
separation $d$ (panel a; $L=1.2\,\lf$, $\Me=50$): it is largest,
$1.64\,\%$ (CI $[1.39,1.90]\,\%$), at $d=1.53\,\lf$, adjacent to the
chosen value $d=1.5\,\lf$, with smaller maxima of approximately
$1.0\,\%$ near $d\approx1.9$ and $2.9\,\lf$. As a function of the
memory (panel b; $d=1.5\,\lf$, $L=1.2\,\lf$), it rises
non-monotonically from near zero at $\Me<10$, passes through a local
maximum near $\Me=50$ and reaches $2.41\,\%$ (CI $[2.12,2.71]\,\%$) at
$\Me\approx132$, near the upper end of the range examined. As a
function of the slit width (panel c; $d=1.5\,\lf$, $\Me=50$), it is
largest, $1.71\,\%$ (CI $[1.46,1.97]\,\%$), at $L=1.17\,\lf$, adjacent
to the chosen value $L=1.2\,\lf$, and vanishes for
$L\lesssim0.5\,\lf$.

\subsection*{S4. Dependence on the kernel decay length}

\noindent
Table~\ref{tab:delta} reports the loop fraction of the main text
remeasured at six values of the decay length $\delta$ of the
single-impact kernel $\hzero(r)=J_0(k_F r)e^{-r/\delta}$, with all
other parameters and the launch seeds held fixed; for each $\delta$
the source-interaction matrix is rebuilt and the wave-force amplitude
$\alpha$ recalibrated to the fixed walking speed $V_0$ (Methods). The
loop fraction collapses for $\delta\leq1.2\,\lf$, where the kernel
does not span adjacent slits ($d=1.5\,\lf$), and grows through
$\delta=2.0$--$2.5\,\lf$. The population of walkers that reach the
screen shrinks rapidly beyond $\delta=2\,\lf$ as the interaction with
the barrier intensifies, which degrades the statistics at
$\delta=3\,\lf$.

\begin{table}[h]
\caption{\textbf{Loop fraction versus kernel decay length.}
Headline protocol at the chosen geometry ($d=1.5\,\lf$, $L=1.2\,\lf$,
$\Me=50$): three-slit loops of walkers launched through $S_3$, counted
among screen arrivals at $y_s=10\,\lf$; $N=20{,}000$ launches per row with
identical seeds; 95\,\% bootstrap CIs. The $\delta=1.5\,\lf$ row is
the headline measurement of the main text.}
\label{tab:delta}
\small
\centering
\begin{tabular}{ccccc}
\toprule
$\delta/\lf$ & screen arrivals & loopers & $f_{\mathrm{loop},3}$ (\%) & 95\,\% CI \\
\midrule
$1.0$ & $14{,}252$ & $8$   & $0.06$ & $[0.02, 0.10]$ \\
$1.2$ & $14{,}538$ & $10$  & $0.07$ & $[0.03, 0.12]$ \\
\rowcolor{green!10}
$\mathbf{1.5}$ & $13{,}334$ & $107$ & $\mathbf{0.80}$ & $[0.65, 0.96]$ \\
$2.0$ & $4{,}909$  & $48$  & $0.98$ & $[0.71, 1.28]$ \\
$2.5$ & $846$      & $27$  & $3.19$ & $[2.01, 4.37]$ \\
$3.0$ & $283$      & $6$   & $2.12$ & $[0.71, 3.89]$ \\
\bottomrule
\end{tabular}
\end{table}

\subsection*{S5. Comparison with quantum backflow}

\noindent
We now compare the loop trajectories of the walker with the analogous
behaviour in the corresponding quantum system. If we impose
isometry between the two systems, can we obtain comparable statistics
--- the likelihood of loops --- in the equivalent quantum system? The quantum-mechanical
description is fundamentally different: there are no trajectories, only
a wavefunction $\psi(\boldsymbol{x},t)$ evolving under the
time-dependent Schr\"{o}dinger equation, whose squared modulus
prescribes the probability density of position measurements. Backward
transport must therefore be diagnosed from the probability current
rather than from particle paths. We integrate the two-dimensional
Schr\"{o}dinger equation for a Gaussian wavepacket launched in the
$+\hat{y}$ direction towards a barrier isometric to that of
Fig.~\ref{fig:diffraction}, realized as a thin rectangular potential
barrier (see `Schr\"{o}dinger reference' below). The dynamics of
Schr\"{o}dinger-type wavepackets has also been studied experimentally
in a hydrodynamic setting, with surface gravity
waves~\citep{Rozenman2020}.

The analogous roles of the Faraday wavelength $\lf$ in pilot-wave
hydrodynamics and the de~Broglie wavelength
$\lambda_{\mathrm{dB}}=2\pi/k_0$ in quantum mechanics are well
established~\citep{CouderFort2006,Bush2015AnnRev,BushOza2021};
isometry thus determines the slit geometry,
$(d,L)=(1.5,\,1.2)\,\lambda_{\mathrm{dB}}$, and sets
$\lambda_{\mathrm{dB}}=\lf$ ($k_0=k_F$) for the baseline packet; the
remaining wavepacket parameters are free. We also employ a
\emph{broadened} packet ($k_0=\pi/2$, $\sigma_y=0.223\,\lf$), which we
refer to as the matched packet: backflow is carried by momentum
components near $k_y\approx0^{+}$, and broadening the packet moves
spectral weight into this near-zero tail. Its parameters are selected,
not derived. The choice is motivated by the wave field that produces
the loops: the spatial power spectrum of the looper profile of
Supplementary Fig.~\ref{fig:vprofiles} comprises a dominant component
at the Faraday wavenumber $k_F$ and a weaker low-wavenumber envelope
peaking near $k\approx2\,\lf^{-1}$, which the matched packet overlaps
and the baseline packet does not.

The observable is the backward probability flux at the slit plane,
\begin{equation}
\Pback
 = \int_{0}^{T}\!\!dt\!\!\int_{\mathrm{slits}}\!\!dx\;
   \max\!\bigl[-j_y(x,0,t),\,0\bigr],
\label{eq:backflow}
\end{equation}
where $j_y = \mathrm{Im}[\psi^*\partial_y\psi]$ is the $y$-component of
the probability current. For a classical positive-momentum ensemble
this integrand vanishes identically; for a quantum state built entirely
from positive-momentum eigenstates it need not. We henceforth refer to the integral of
Eq.~\eqref{eq:backflow} for a general packet as the \emph{retrograde
flux}; only for states with strictly non-negative momentum content
does it constitute quantum backflow
proper~\citep{Bostelmann2017}.

\begin{figure}[tbp]
\centering
\includegraphics[width=0.93\linewidth]{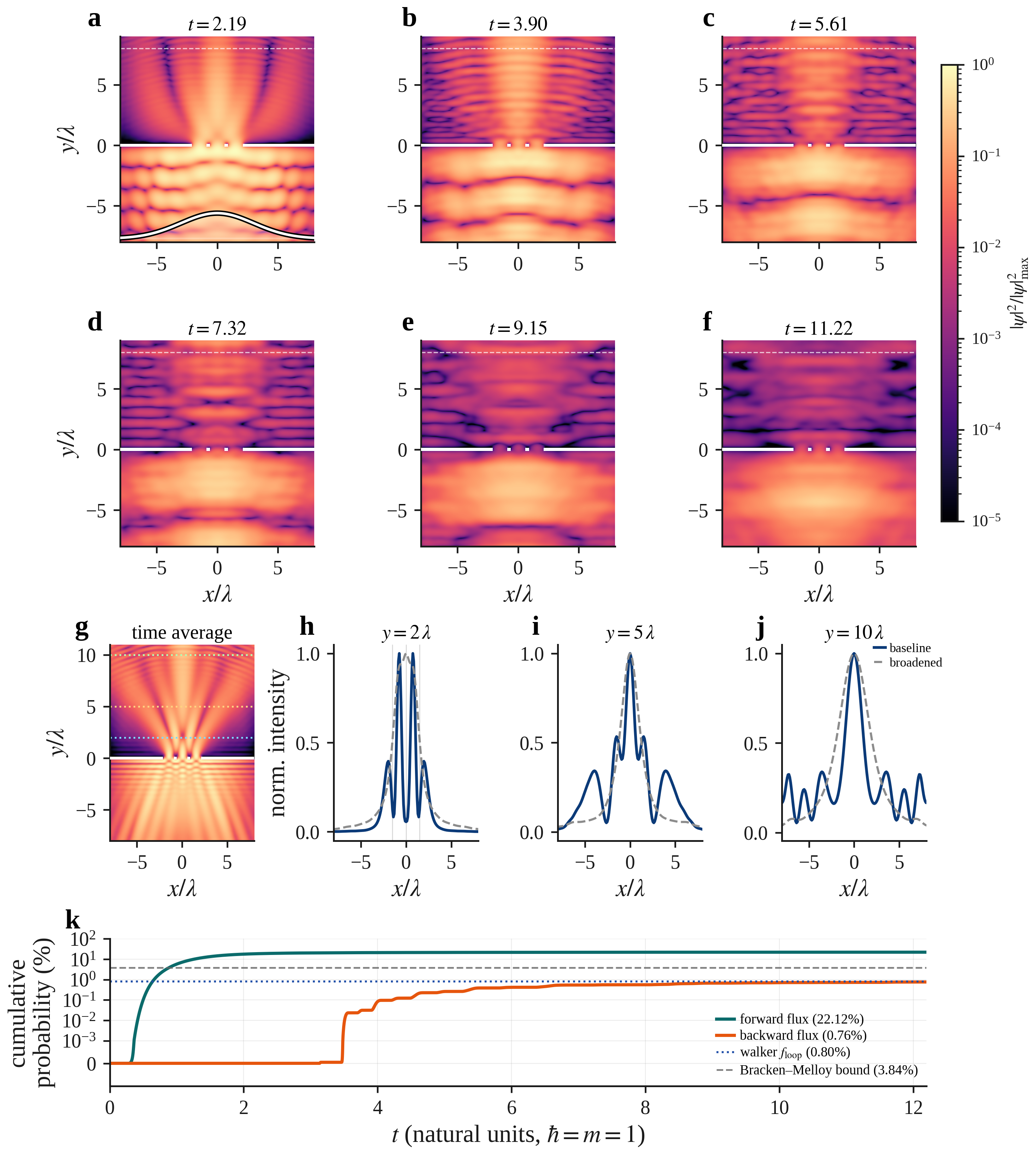}
\caption{\textbf{Backward probability flux of a Schr\"{o}dinger packet
through the triple-slit geometry.}
The evolution of a Gaussian wavepacket launched towards the slits
with positive mean momentum.
(a--f)~$|\psi(x,y,t)|^2$ at six successive times (shared logarithmic
colour scale, normalized to the global peak) for the broadened packet
[$\sigma_x=3.0\,\lf$, $\sigma_y=0.223\,\lf$, $k_0=\pi/2$]; the white curve along the base
of (a) shows the initial profile $|\psi_0(x)|^2$, a Gaussian packet
elongated along $x$. Even though the packet moves forward, a small part of the
probability current at the slit plane runs backward.
(g)~Time-averaged density $\langle|\psi|^2\rangle$ of the
baseline packet ($k_0=k_F$), the quantum counterpart of the walker
density maps of Fig.~\ref{fig:datagrid}d--f; dotted lines mark the
near-field ($y=2\,\lambda$), mid-field ($y=5\,\lambda$) and
far-field ($y=10\,\lambda$) cuts, with lengths in this figure
expressed in units of the common wavelength
$\lambda=\lambda_{\mathrm{dB}}=\lf$.
(h--j)~Intensity profiles at $y=2\,\lambda$, $5\,\lambda$ and
$10\,\lambda$: the baseline packet (solid) develops the three-slit
interference structure (slit positions in grey in h) and, further
downstream, a multi-lobed pattern, whereas for the broadened packet
(dashed, $\lambda_{\mathrm{dB}}=4\,\lf>d$) all orders beyond the
zeroth are evanescent and the profile remains a single lobe.
(k)~Cumulative fluxes from Eq.~\eqref{eq:backflow}: forward (teal,
saturating near $22\,\%$) and backward (orange, $0.76\,\%$), with the
walker loop fraction (blue dotted, $0.80\,\%$) and the Bracken--Melloy
bound (grey dashed, $3.84\,\%$).}
\label{fig:backflow}
\end{figure}

Supplementary Figure~\ref{fig:backflow} shows the transit of the
matched packet ($\sigma_x=3.0\,\lf$) together with the diffraction
patterns of the baseline packet, the quantum counterpart of the walker
patterns of Fig.~\ref{fig:datagrid}; for the matched packet
($\lambda_{\mathrm{dB}}=4\,\lf>d$) all orders beyond the zeroth are
evanescent, and the transmitted density is a single smooth lobe. The
matched packet yields a forward transmission $\Pfwd=22.12\,\%$ and a
retrograde flux $\Pback=0.76\,\%$, a factor $\sim\!620$ above that of
the nearly monochromatic baseline ($\sigma_k/k_0\approx0.05$;
$1.23\times10^{-3}\,\%$;
Table~\ref{tab:sweep})~\citep{BrackenMelloy1994,Penz2006,Yearsley2012,Halliwell2013}.
Because the matched packet carries $22.4\,\%$ of its momentum density
at $k_y<0$, part of its retrograde flux is classically allowed.
Repeating the calculation with the packet projected onto $k_y\geq0$ at
$t=0$ (see below) yields $\Pback^{+}=0.86\,\%$: for such a state
traversing a thin barrier, no classical particle with positive
momentum can recross the slit plane, so all of $\Pback^{+}$ is
genuine, classically forbidden backflow. The evolution of the matched
packet is shown in Supplementary Fig.~\ref{fig:qmatched} for
comparison with the walker trajectories of
Fig.~\ref{fig:diffraction}c.

Nor are the initial conditions of the two systems fully analogous. The
walkers are launched through $S_3$ with impact parameters distributed
uniformly across the slit (Sec.~S1), whereas the packet is launched on
the centreline with a Gaussian transverse profile of width
$\sigma_x=3.0\,\lf$, whose density varies by $22\,\%$ across the slit
array. To assess the influence of the transverse profile, we repeated
the projected calculation with flat-topped profiles, for which the
transverse density $|\psi_0(x)|^2$ is uniform over a prescribed range,
its edges being smoothed over $0.2\,\lf$ (Table~\ref{tab:sweep}).
Uniform illumination of the slit array ($|x|\leq2.1\,\lf$) yields
$\Pback^{+}=0.55\,\%$; wider profiles ($|x|\leq4$ and $6\,\lf$) yield
$0.96\,\%$ and $0.95\,\%$; and a profile that spans $S_3$ alone at
launch, the counterpart of the walker ensemble of the main text,
yields $0.32\,\%$. The last does not remain confined to $S_3$: as
$\lambda_{\mathrm{dB}}=4\,\lf$ exceeds the slit width, the packet
spreads before reaching the barrier, where only about one fifth of its
density faces $S_3$. These values change by at most $0.05\,\%$ as the
width of the smoothed edges is varied between $0.1$ and $0.3\,\lf$.
The backflow thus lies between $0.3\,\%$ and $1.0\,\%$ for all the
profiles considered: its order of magnitude, though not its precise
value, is independent of the transverse profile. The evolution of
the widest of these packets is shown in Supplementary
Fig.~\ref{fig:qflat}; before projection, its retrograde flux is
$\Pback=0.86\,\%$, against $0.76\,\%$ for the Gaussian packet. For
walkers launched uniformly across all three slits, as the packet
illuminates them, the grids of Secs.~S1 and S2 give a loop fraction of
$0.63\,\%$ among screen arrivals and $0.41\,\%$ among launches.

\emph{The compared observables are distinct.} $f_{\mathrm{loop},3}$ is a
trajectory-classification probability, whereas $\Pback$ is an
integrated negative-current flux through the aperture plane; the walker
follows a deterministic trajectory in a classical wave field, while the
quantum packet has no trajectories at all. The two measures need not
coincide, and the photonic benchmark of
Ref.~\citep{MaganaLoaiza2016} likewise compares non-identical
observables, inferring loop trajectories from far-field statistics. The
comparison should therefore be read at the level of scale: the
backward transport amounts to between a few tenths of one per cent and
one per cent in both systems (see below for normalization
conventions).

\begin{figure}[tbp]
\centering
\includegraphics[width=0.96\linewidth]{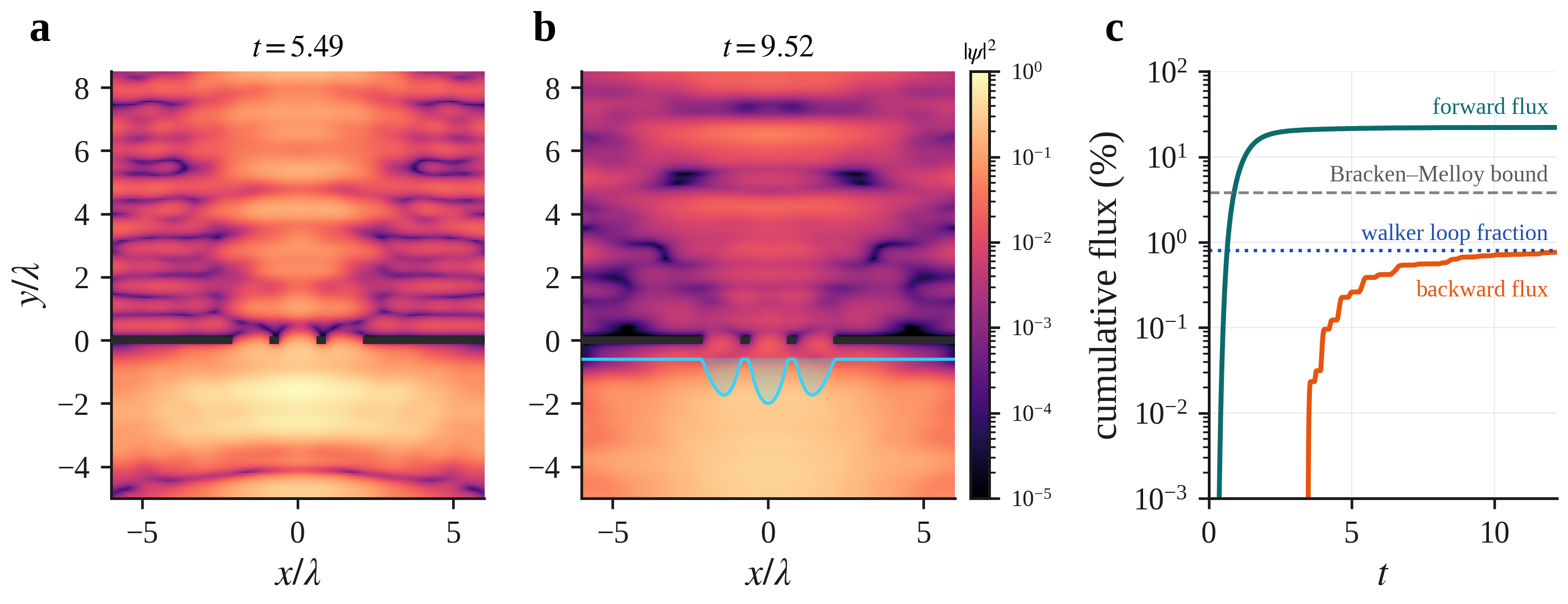}
\caption{\textbf{Evolution of the wavepacket matched to the walker's
spectral envelope.}
(a)~Matched wavepacket ($k_0=\pi/2$, $\sigma_y=0.223\,\lf$) at $t=5.49$
($|\psi|^2$ normalized to its peak, logarithmic colour scale),
traversing the barrier isometric to that of
Fig.~\ref{fig:diffraction}.
(b)~$|\psi|^2$ at $t=9.52$ with the accumulated retrograde current
density $\int_0^t dt'\max[-j_y,0]$ overlaid (cyan).
(c)~Cumulative forward (teal) and retrograde (orange) fluxes, with the
walker loop fraction $f_{\mathrm{loop},3}=0.80\,\%$ (blue dotted) and
the Bracken--Melloy bound ($3.84\,\%$, grey dashed).}
\label{fig:qmatched}
\end{figure}

\begin{figure}[tbp]
\centering
\includegraphics[width=0.96\linewidth]{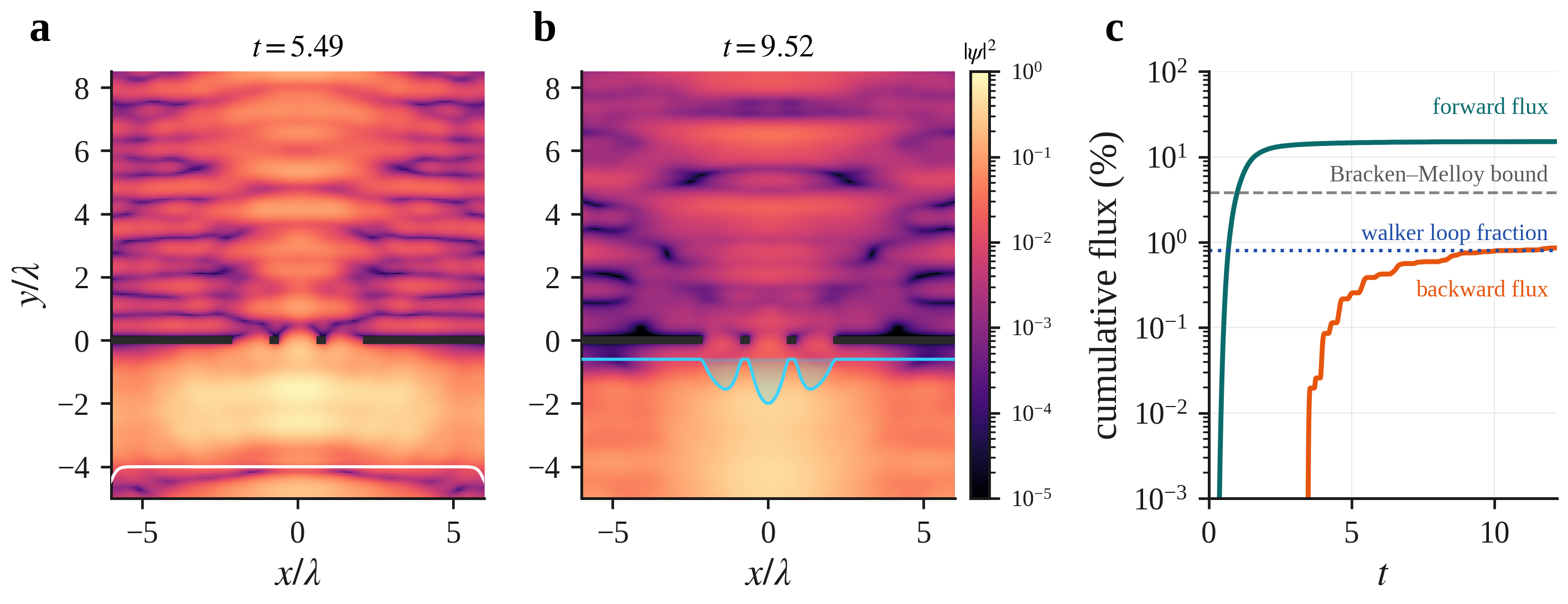}
\caption{\textbf{Evolution of a wavepacket with a flat-topped
transverse profile.}
As Supplementary Fig.~\ref{fig:qmatched}, for the packet whose
transverse density is uniform over $|x|\leq6\,\lf$ (white curve at
the base of a), with $k_0=\pi/2$ and $\sigma_y=0.223\,\lf$.
(a)~$|\psi|^2$ at $t=5.49$.
(b)~$|\psi|^2$ at $t=9.52$ with the accumulated retrograde current
density overlaid (cyan).
(c)~Cumulative forward (teal, $\Pfwd=15.10\,\%$) and retrograde
(orange, $\Pback=0.86\,\%$) fluxes, with the walker loop fraction
$f_{\mathrm{loop},3}=0.80\,\%$ (blue dotted) and the Bracken--Melloy
bound ($3.84\,\%$, grey dashed). The $k_y\geq0$ projection of this
packet yields $\Pback^{+}=0.95\,\%$ (Table~\ref{tab:sweep}).}
\label{fig:qflat}
\end{figure}

\begin{table}[h]
\caption{\textbf{Dependence on wavepacket parameters:
Schr\"{o}dinger retrograde flux
and backflow through the three-slit barrier.}
All runs: barrier isometric to that of Fig.~\ref{fig:diffraction}
($d=1.5\,\lf$, $L=1.2\,\lf$, $V_{\max}=400$), wavepacket launched at
$(0,-6\,\lf)$ except the matched and flat-topped rows (launched at
$(0,-4\,\lf)$).
$\Pback$ is evaluated from Eq.~\eqref{eq:backflow}; $\Pfwd$ is the
cumulative positive flux through the apertures. The matched row is the
wavepacket of Figs.~\ref{fig:backflow} and~\ref{fig:qmatched}; the
row beneath it is its strict $k_y\geq0$ projection
($^{\dagger}$Gaussian parameters before projection; see below).
In the last four rows the Gaussian transverse profile of the matched
packet is replaced by a flat-topped one, uniform over the range
indicated with edges smoothed over $0.2\,\lf$, and the packet is
likewise projected onto $k_y\geq0$.}
\label{tab:sweep}
\small
\centering
\begin{tabular}{lcccccc}
\toprule
Configuration & $k_0$ & $\sigma_y$ & $y_0$ &
  $\Pfwd$ (\%) & $\Pback$ (\%) & $\Pback/\Pfwd$ \\
\midrule
Baseline (narrow)
  & $2\pi$ & $1.5$ & $-6$ & $39.21$ & $1.23\times10^{-3}$ & $3.1\times10^{-5}$ \\
Slower
  & $\pi$  & $1.5$ & $-6$ & $48.00$ & $1.43\times10^{-3}$ & $3.0\times10^{-5}$ \\
Much slower
  & $\pi/2$& $1.5$ & $-6$ & $16.97$ & $9.56\times10^{-3}$ & $5.6\times10^{-4}$ \\
Narrower
  & $\pi$  & $0.7$ & $-6$ & $44.64$ & $4.78\times10^{-2}$ & $1.1\times10^{-3}$ \\
Both
  & $\pi/2$& $0.7$ & $-6$ & $18.46$ & $1.16\times10^{-1}$ & $6.3\times10^{-3}$ \\
\rowcolor{green!10}
\textbf{Matched}
  & $\pi/2$& $0.223$ & $-4$ & $22.12$ & $\mathbf{0.762}$   & $3.4\times10^{-2}$ \\
\rowcolor{green!20}
\textbf{Matched, $k_y\!\geq\!0$ projected}
  & $\pi/2$& $0.223^{\dagger}$ & $-4$ & $28.41$ & $\mathbf{0.856}$   & $3.0\times10^{-2}$ \\
\midrule
\multicolumn{7}{l}{\emph{Flat-topped transverse profile, $k_y\!\geq\!0$ projected}}\\
Slit array, $|x|\leq2.1\,\lf$
  & $\pi/2$& $0.223^{\dagger}$ & $-4$ & $45.45$ & $0.550$ & $1.2\times10^{-2}$ \\
$|x|\leq4\,\lf$
  & $\pi/2$& $0.223^{\dagger}$ & $-4$ & $32.81$ & $0.955$ & $2.9\times10^{-2}$ \\
$|x|\leq6\,\lf$
  & $\pi/2$& $0.223^{\dagger}$ & $-4$ & $19.36$ & $0.951$ & $4.9\times10^{-2}$ \\
$S_3$ alone, $0.9\leq x\leq2.1\,\lf$
  & $\pi/2$& $0.223^{\dagger}$ & $-4$ & $31.88$ & $0.320$ & $1.0\times10^{-2}$ \\
\bottomrule
\end{tabular}
\end{table}

\paragraph{Schr\"{o}dinger reference.}
The 2D time-dependent Schr\"{o}dinger equation
($\hbar=m=1$, split-step Fourier, $256\times 256$ grid,
$L_x=L_y=24\,\lf$, $\Delta t=10^{-3}$; $\Delta t=2\times10^{-3}$ for
the matched and flat-topped packets, for which halving the step changes
$\Pback$ by no more than $0.001\,\%$) is integrated with a
rectangular barrier $V_{\max}=400$ of thickness $0.25\,\lf$ at $y=0$,
leaving slit apertures open. An imaginary absorbing potential of
strength $5$ over a $1.5\,\lf$ outer shell prevents wrap-around
reflections. The probability current $j_y$ is evaluated by central
differences at $y=0$ and integrated according to
Eq.~\eqref{eq:backflow}. The initial wavepacket is
\begin{equation}
\psi(\boldsymbol{x},0) \propto
  \exp\!\Bigl[-\tfrac{(x-x_0)^2}{4\sigma_x^2}
              -\tfrac{(y-y_0)^2}{4\sigma_y^2}
              + i k_0 (y-y_0)\Bigr],
\label{eq:psi0}
\end{equation}
so that $\sigma_y$ is the standard deviation of the position density
$|\psi|^2$ and its Fourier transform
$\tilde\psi(k_y)\propto\exp[-\sigma_y^2(k_y-k_0)^2]$ has momentum
density $|\tilde\psi|^2\propto\exp[-(k_y-k_0)^2/2\sigma_k^2]$ with
\emph{momentum standard deviation} $\sigma_k=1/(2\sigma_y)$; we report
the relative bandwidth as $\sigma_k/k_0$ throughout. The wavepacket parameters in
Table~\ref{tab:sweep} vary only $k_0$, $\sigma_y$ and launch
position $y_0$, with $\sigma_x=3.0\,\lf$ throughout. For the
flat-topped profiles, the Gaussian factor in $x$ is replaced by
$\{[\tanh((x-x_a)/w)-\tanh((x-x_b)/w)]/2\}^{1/2}$ with $w=0.2\,\lf$,
so that $|\psi|^2$ is uniform between $x_a$ and $x_b$. The matched
configuration uses $k_0=\pi/2$, $\sigma_y=0.223\,\lf$, $y_0=-4\,\lf$,
giving $\sigma_k/k_0\approx 1.4$. Consistent with this width, a
fraction $22.4\,\%$ of the packet's momentum density lies at $k_y<0$
(the value expected from Eq.~\eqref{eq:psi0} for $\sigma_k=1/(2\sigma_y)$
is $24\,\%$); the associated classically-allowed reverse current is
included in $\Pback$ as defined by Eq.~\eqref{eq:backflow}.

\paragraph{Positive-momentum projection.}
The strict-backflow reference is constructed by projecting the
matched packet onto non-negative momenta $k_y$ at $t=0$:
$\tilde\psi(k_x,k_y)\mapsto 0$ for $k_y<0$ (the negative-signed
Nyquist bin included), followed by renormalization. The projection
retains $77.6\,\%$ of the norm and leaves a residual $k_y<0$ fraction
below $10^{-29}$. Because strictly positive-momentum states cannot
vanish on a half-line, the projected packet carries an unavoidable
algebraic tail: at $t=0$ it places $0.80\,\%$ of the probability
beyond the barrier rear face, which we quote as a bound on any
spurious early-time flux contribution. Evolving with the identical
solver gives $\Pback^{+}=0.856\,\%$, $\Pfwd=28.4\,\%$ at
$\Delta t=2\times10^{-3}$ and $0.857\,\%$, $28.5\,\%$ at
$\Delta t=10^{-3}$; the unprojected matched run is likewise
step-converged ($\Pback=0.762\,\%$ at both $\Delta t$). The
backflow so obtained is the scattering-geometry counterpart of
free-state Bracken--Melloy backflow~\cite{Bostelmann2017}.

\paragraph{Normalization conventions.} $f_{\mathrm{loop},3}$ is quoted per
walker arriving at the screen, and $\Pback$ per total initial norm. Under
parallel normalizations the agreement between the two systems is at
the order-of-magnitude level: per launched event the walker rate is
$0.54\,\%$ against the total-norm backflow $0.86\,\%$ (a factor
$\sim\!1.6$); conditioned on transmission, the walker gives
$0.80\,\%$ of screen arrivals against
$\Pback^{+}/\Pfwd\approx3\,\%$ of forward flux (a factor
$\sim\!4$).

\end{document}